\documentclass[amsmath,amssymb,aps,prx,reprint,superscriptaddress]{revtex4-1}
\usepackage{graphicx}% Include figure files
\usepackage{xcolor}
\usepackage[colorlinks=true,urlcolor=blue,citecolor=blue,linkcolor=blue]{hyperref}
\usepackage{physics}
\usepackage{siunitx}
\usepackage{xspace}
\usepackage{bm}
\usepackage{amssymb,ulem,amsmath}
\usepackage{verbatim}
\usepackage{comment}
\newcommand{\vect}[1]{\vb{#1}}
\newcommand{\transpose}{\mathsf{T}}

\newcommand{\bk}{\textbf{k}}
\newcommand{\bq}{\textbf{q}}

\newcommand{\br}{\vect{r}}

\definecolor{AJ-color}{named}{blue}
\definecolor{PT-color}{named}{orange}
\definecolor{GS-color}{named}{purple}

\begin{document}

\title{Supefluidity of flat band Bose-Einstein condensates revisited}

\author{Aleksi Julku}
\email{ajulku@phys.au.dk}
\affiliation{\affiliationAarhus}
\author{Grazia Salerno}
\affiliation{\affiliationAalto}
\author{P\"aivi T\"orm\"a}
\email{paivi.torma@aalto.fi}
\affiliation{\affiliationAalto}
\newcommand{\affiliationAalto}{Department of Applied Physics, Aalto University, P.O.Box 15100, 00076 Aalto, Finland}
\newcommand{\affiliationAarhus}{Center for Complex Quantum Systems, Department of Physics and Astronomy, Aarhus University, Ny Munkegade 120, DK-8000 Aarhus C, Denmark}
\newcommand{\affiliationChina}{Shenzhen Institute for Quantum Science and Engineering and Department of Physics, Southern University of Science and Technology, Shenzhen 518055, China}


\date{\today}

\begin{abstract}
We consider the superfluid weight, speed of sound and excitation fraction of a flat band Bose-Einstein condensate (BEC) within multiband Bogoliubov theory. The superfluid weight is calculated by introducing a phase winding and minimizing the free energy with respect to it. We find that the superfluid weight has a contribution arising from the change of the condensate density and chemical potential upon the phase twist that has been neglected in the previous literature. We also point out that the speed of sound and the excitation fraction are proportional to orbital-position-independent generalizations of the quantum metric and the quantum distance, and reduce to the usual quantum metric (Fubini-Study metric) and the Hilbert-Schmidt quantum distance only in special cases. We derive a second order perturbation correction to the dependence of the speed of sound on the generalized quantum metric, and show that it compares well with numerical calculations. Our results provide a consistent connection between flat band BEC and quantum geometry, with physical observables being independent of the orbital positions for fixed hopping amplitudes, as they should, and complete formulas for the evaluation of the superfluid weight within the Bogoliubov theory. We discuss the limitations of the Bogoliubov theory in evaluating the superfluid weight.       
        
\end{abstract}

%\keywords{}

\maketitle

\section{Introduction}

Flat (dispersionless) bands are of interest since interaction effects dominate over kinetic energy and interaction-driven phenomena such as superconductivity or ferromagnetism may become enhanced. Furthermore, as recent work has shown, \textit{quantum geometry} often plays an interesting role in flat band systems. For instance in flat band superconductivity~\cite{kopnin:2011,heikkila:2011,Peotta2015,Liang2017,Torma2018}, an obvious question is how can a finite supercurrent exist with the electron effective mass being infinite in a flat band? This dilemma was solved in Ref.~\cite{Peotta2015} by proving that the superfluid weight $D^s$ can be divided to single-band ($D^s_{\text{conv}}$) and multiband contributions ($D^s_{\text{geom}}$) so that $D^s = D^s_{\text{conv}} +  D^s_{\text{geom}}$. The multiband contribution can be non-zero also in a flat band; in other words, the Cooper pair mass can be finite despite the infinite effective mass of the electrons~\cite{Torma2018}. In Ref.~\cite{Peotta2015,Liang2017} the multiband contribution was shown to be connected to quantum geometric concepts such as the quantum geometric tensor, quantum metric, and Berry curvature~\cite{Provost:1980,Resta2011}. Furthermore, Bose-Einstein condensates (BECs) have been predicted in flat bands~\cite{Huber2010,you:2012}, and quantum geometry turns out to be crucial for the stability and superfluidity of BECs as well~\cite{Julku2021a,Julku2021b}. Recently, it has been however pointed out that the results on multiband superfluid weight $D^s$ of fermionic superconductivity have been incomplete~\cite{Huhtinen2022,Herzog-Arbeitman2022}. In this article we solve a similar caveat related to multiband, especially flat band, superfluidity of BECs.

Superfluid weight is defined as the change of free energy $F$ upon a phase twist (supercurrent) $q$ introduced to the system: $D^s \propto d^2 F/d q^2 |_{q=0,N_{tot}}$, where the total particle number $N_{tot}$ is kept constant. In the presence of time-reversal symmetry (TRS), this reduces to $D^s \propto d^2 \Omega/d q^2 |_{q=0,N_{tot}}$, with $\Omega$ being the grand potential. If one further assumes, as was done in Ref.~\cite{Peotta2015} where the geometric contribution of superconductivity was for the first time identified, that the superconducting order parameter at each orbital is always real, then   
 $D^s_{\mu\nu} \propto \frac{\partial^2 \Omega}{\partial q_\mu \partial q_\nu} |_{q=0,\mu}$. In Ref.~\cite{Liang2017}, $D^s$ was computed by using the linear response theory which turned out to coincide with the result of  Ref.~\cite{Peotta2015}. From $D^s_{\mu\nu} \propto \frac{\partial^2 \Omega}{\partial q_\mu \partial q_\nu} |_{q=0,\mu}$, it was derived that the superfluid weight in a flat band is proportional to the quantum metric integrated over the momentum space and bounded from below by the Chern number~\cite{Peotta2015} and Berry curvature~\cite{Liang2017}. There remained a problem, though, that the superfluid weight is by definition independent of the positions of the orbitals within the unit cell (assuming the connectivities, i.e., hoppings are fixed), while quantum metric and Berry curvature depend on them. 

In a recent study~\cite{Huhtinen2022}, it was shown that the expression $D^s_{\mu\nu} \propto \frac{\partial^2 \Omega}{\partial q_\mu \partial q_\nu} |_{q=0,\mu}$ for fermionic superconductivity is incomplete within the mean-field theory and one has to carefully take into account the derivatives of the superconducting order parameter. These new terms were shown to always decrease the superfluid weight and in some cases qualitative differences between the old results and the new complete superfluid weight expressions could be revealed. This finding solved the dilemma related to the orbital positions, as it was shown that the superfluid weight is actually proportional to the \textit{minimal quantum metric} (quantum metric with a minimal trace), a quantity that does not depend on orbital positions for fixed hopping parameters (i.e.~does not depend on the orbitals' embedding in the Hamiltonian). Furthemore, when the orbitals are at high-symmetry positions, the quantum metric is automatically minimal. The new terms appear only in multiband systems and thus the new revelation does not affect the usual single-band BCS superconductors.

The new superfluid terms pointed out in Ref.~\cite{Huhtinen2022} were derived for fermionic systems. However, the question remains whether similar arguments apply to bosonic multiband systems as well. In Refs.~\cite{Julku2021a,Julku2021b}, BEC speed of sound and excitation fraction (quantum depletion) were shown to be proportional to the quantum metric and the quantum distance, respectively, and the superfluid weight was computed for bosonic kagome model with the linear response theory that coincides with the old expression $D^s_{\mu\nu} \propto \frac{\partial^2 \Omega}{\partial q_\mu \partial q_\nu} |_{q=0,\mu}$. Thus it is not completely clear whether the superfluid weight calculations of Ref.~\cite{Julku2021b} are complete or not, and whether the connection of the BEC properties to quantum geometric quantities in~\cite{Julku2021a,Julku2021b} needs further inspection. In this article we answer these questions. It turns out that indeed the superfluid weight in the bosonic case also has correction terms that have been omitted in previous literature, originating from the dependence of the condensate density $n_0$ and chemical potential $\mu$ on the phase twist. These are important to consider in the general case, although we show that their effect in the examples studied in~\cite{Julku2021a,Julku2021b} is very small within the Bogoliubov theory. We also discuss the limitations of the Bogoliubov approach. 

We moreover point out that one should be cautious when evaluating the speed of sound and excitation fraction. They are shown in~\cite{Julku2021b} to be related to generalized versions of the quantum metric and the quantum distance; we discuss here that these quantities are independent on the orbital positions for fixed hopping parameters, as they should. On the other hand, we emphasize that the connection to the usual quantum metric and the Hilbert-Schmidt quantum distance -- which do depend on orbital positions -- is valid only under specific conditions for the Bloch functions.

In Section~\ref{theoframe}, the theoretical framework of multiband Bogoliubov theory is introduced. Then, the relation of the speed of sound and the excitation fraction to the quantum metric and distance is considered in Section~\ref{metricsection}. We then compute the superfluid weight in Section~\ref{sfwsection}. In Section~\ref{Sec_cs} we furthermore provide the second order perturbation result for the speed of sound in case of the flat band BEC. We finally conclude in Section~\ref{conclusions}.

\section{Theoretical framework of multiband BEC}  \label{theoframe}

We start by considering a generic multiband Bose-Hubbard grand-canonical Hamiltonian 
\begin{align}
\label{BHH}
&H = H_0 + H_{\text{int}} - \mu \sum_{i\alpha} c^\dag_{i\alpha} c_{i\alpha}, \quad\text{with} \\
& H_0 = \sum_{i\alpha j\beta} t_{i\alpha j\beta}c^\dag_{i\alpha} c_{j\beta}\\
& H_{\text{int}} = \frac{U}{2}\sum_{i\alpha} c^\dag_{i\alpha} c^\dag_{i\alpha} c_{i\alpha} c_{i\alpha}.
\end{align}
Here $H_0$ is the kinetic and $H_{\text{int}}$ is the interaction Hamiltonian. Moreover, $c_{i\alpha}$ is a bosonic annihilation operator for the $\alpha$th sublattice (in the following, we use terms orbital and sublattice interchangeably) within the $i$th unit cell, $t_{i\alpha,j\beta}$ is the kinetic hopping term, $\mu$ is the chemical potential and $U>0$ is repulsive on-site interaction. The sublattice index $\alpha$ runs from $1$ to $M$, where $M$ is the number of lattice sites per unit cell (i.e the number of sublattices). For example in a honeycomb lattice one has $M=2$ and kagome geometry $M=3$. One can introduce the Fourier transform
\begin{align}
c_{i\alpha} = \frac{1}{\sqrt{N}}\sum_\bk e^{i \bk \cdot \br_{i\alpha}} c_{\bk\alpha}, 
\end{align}
where $N$ is the number of unit cells and $\br_{i\alpha} = \br_i + \br_\alpha$ with $\br_i$ being the spatial coordinate of unit cell $i$ and $\br_\alpha$ the coordinate of a site belonging to sublattice $\alpha$ within a unit cell. One can then recast the non-interacting Hamiltonian $H_0$ as 
\begin{align}
 H_0 = \sum_{\bk} c^\dag_{\bk\alpha}\mathcal{H}_{\alpha\beta}(\bk)c_{\bk\beta} \equiv \sum_\bk c^\dag_\bk \mathcal{H}(\bk)c_\bk,   
\end{align}
where $\mathcal{H}(\bk)$ is a $M\times M$ matrix and $c_\bk$ is a $M\times 1$ vector such that  $[\mathcal{H}(\bk)]_{\alpha\beta} = \mathcal{H}_{\alpha\beta}(\bk)$ and $[c_\bk]_\alpha = c_{\bk\alpha}$. The hopping elements in the momentum space read 
\begin{align}
\label{hopping_element}
 \mathcal{H}_{\alpha\beta}(\bk) = \sum_{\br_{\alpha\beta}} t(\br_{\alpha\beta}) e^{-i\br_{\alpha\beta}\cdot \bk},   
\end{align}
where $\br_{\alpha\beta}$ are all possible vectors connecting a lattice site of  sublattice index $\alpha$ to sites residing in sublattice $\beta$ and $t(\br_{\alpha\beta})$ are the corresponding hopping terms. This form follows from the translational invariance of the hopping Hamiltonian.

One can diagonalize $\mathcal{H}(\bk)$ as $\mathcal{H}(\bk) |u_n(\bk) \rangle = \epsilon_n(\bk)|u_{n}(\bk) \rangle$, where $\epsilon_n(\bk)$ are the eigenenergies and $|u_{n\bk} \rangle$ are the corresponding periodic parts of the Bloch states ($n$ is the band index). The Bloch band energies are ordered in the ascending order, i.e. $\epsilon_1(\bk) \leq \epsilon_2(\bk) \leq ... \leq \epsilon_M(\bk)$ for all $\bk$. Explicitly, one has
\begin{align}
&H_0 = \sum_\bk c^\dag_\bk U(\bk) D(\bk) U^\dag(\bk) c_\bk \equiv \sum_\bk \gamma^\dag_\bk D(\bk) \gamma_\bk.  
\end{align}
Here $D(\bk)$ is a diagonal matrix containing the energies of the Bloch bands, i.e. $[D(\bk)]_{nn} = \epsilon_n(\bk)$, and the columns of $U(\bk)$ contain the corresponding Bloch functions, i.e. $[U(\bk)]_{\alpha n} = \langle \alpha | u_n(\bk) \rangle$. The annihilation operators for the Bloch states are expressed as $\gamma_\bk = U^\dag(\bk) c_\bk$ so that $[\gamma_\bk]_n = \gamma_{\bk n}$ where $\gamma_{\bk n}$ is the annihilation operator for the Bloch state of momentum $\bk$ within the $n$th Bloch band. 

As we are dealing with equilibrium physics, it is plausible to assume that Bose-Einstein condensation takes place within the lowest Bloch band at momentum $\bk_c$ so that the bosons condense at the Bloch state $ |\phi_0 \rangle \equiv |u_1(\bk_c)\rangle $ at the energy $\epsilon_0 \equiv \epsilon_1(\bk_c)$ with the corresponding BEC wavefunction being $\psi_0(\br_{i\alpha}) = \exp(i\bk_c \cdot \br_{i\alpha}) \langle \alpha | \phi_0 \rangle$. Note that here we assume that the condensate takes place within a single momentum. If this was not the case, the unit cell can often be expanded such that all the condensate momenta are folded back to a single momentum, i.e. to $\bk = 0$. In the weak-coupling regime, the condensate Bloch state and its momentum $\bk_c$ can be solved by minimizing the corresponding mean-field energy as a function of $\bk_c$~\cite{you:2012,Julku2021b}.

In the case of a dispersive band and weak interaction regime, the condensation can be usually assumed to take place within the Bloch state of the lowest kinetic energy. For example for a square lattice this would mean $\bk_c = 0$. However, here our primary focus is to consider a BEC taking place within a flat band for which all the momentum states have the same kinetic energy and thus the BEC emerges in the Bloch state that minimizes the repulsive on-site interaction energy. Consequently, particles try to distribute as uniformly as possible among all the sublattices~\cite{you:2012,Julku2021a,Julku2021b}. In the following, we thus assume the uniform condensate density condition $|\langle \alpha | \phi_0 \rangle|^2 = \frac{1}{M}$. This is analogous to the uniform pairing condition often assumed in fermionic flat band superconductivity studies~\cite{Peotta2015,Tovmasyan2016}. This condition is important and interesting since it requires a local (point group) symmetry of a specific type to be enforced~\cite{Herzog-Arbeitman2022}. 

We now take into account the quantum fluctuations around the BEC by writing the bosonic annihilation operators as 
\begin{align}
\label{spoint}
c_{i\alpha} = \sqrt{n_0} \psi_0(\br_{i\alpha}) + \delta c_{i\alpha} = \sqrt{n_0}e^{i \bk_c \cdot \br_{i\alpha}} \langle \alpha | \phi_0 \rangle + \delta c_{i\alpha},    
\end{align}
where $n_0$ is the condensation density, i.e. number of condensed bosons per \textit{unit cell} and $\delta c_{i \alpha}$ describes the fluctuations on top of the condensate. By Fourier transforming, one finds 
 $c_{\bk \alpha}= \sqrt{Nn_0}\langle  \alpha | \phi_0 \rangle \delta_{\bk,\bk_c} + \delta c_{\bk\alpha}$ and $\gamma_{\bk n} = \sqrt{Nn_0}\delta_{\bk,\bk_c}\delta_{n,1} + \delta \gamma_{\bk n}$.

As we are considering a system of weakly interacting Bose-condensed gas, we treat the Hamiltonian within the multiband Bogoliubov approximation by neglecting the interaction terms that are higher than quadratic order in the fluctuations $\delta c_{\bk\alpha}$ and $\delta c^\dag_{\bk\alpha}$ with $\bk\neq\bk_c$ and furthermore ignoring the anomalous off-diagonal self-energy contribution. As a result, we write our Bogoliubov Hamiltonian as (for details, see Appendix~\ref{AppA})
\begin{align}
\label{HPopov}
&\frac{H}{N} = (\epsilon_0 - \mu)n_0 + \frac{Un_0^2}{2}\sum_\alpha |\langle \alpha | \phi_0 \rangle|^4 \nonumber \\
&+ \frac{1}{N} \sum_{\bk\alpha\beta} (\mathcal{H}_{\alpha \beta}(\bk) - \mu \delta_{\alpha\beta}) \delta c^\dag_{\bk \alpha} \delta c_{\bk \beta} \nonumber \\
& + \frac{U}{2N} \sum_{\bk \alpha} \Big[ 4n_0 |\langle \alpha | \phi_0 \rangle |^2  \delta c^\dag_{\bk \alpha} \delta c_{\bk \alpha} + \nonumber \\
&n_0 \Big( \langle \alpha | \phi_0 \rangle^2 \delta c^\dag_{\bk \alpha} \delta c^\dag_{2\bk_c-\bk \alpha} + \text{h.c.} \Big) \Big].
\end{align}
Here the first line is constant and describes the condensate while second line is the kinetic energy of the fluctuations and the last two lines are the interaction Hamiltonian written in the Bogoliubov approximation. 

The chemical potential $\mu$ can be solved by demanding that the Hamiltonian terms linearly proportional to the fluctuation of the BEC, i.e. $\delta \gamma_{\bk_c 1}$, vanish. Thus, by plugging Eq.~\eqref{spoint} to~\eqref{BHH} and demanding that linear terms in  $\delta \gamma_{\bk_c 1}$ are zero, one finds (see Appendix~\ref{AppA})
\begin{align}
\label{muexp}
\mu = \epsilon_0 + Un_0 \sum_\alpha |\langle \alpha | \phi_0 \rangle|^4 .
\end{align}
With uniform BEC condition, this reduces to $\mu = \epsilon_0 + \frac{Un_0}{M}$, i.e.~to the expression used in Ref.~\cite{Julku2021a,Julku2021b}. By substituting this to Eq.~\eqref{HPopov}, one obtains $H = H_c + H_B$, where $H_c$ is a constant and $H_B$ is the quantum fluctuation contribution describing the quasi-particle excitations on top of the BEC, i.e.
\begin{align}
\label{ham}
H_B = \frac{1}{2}\sum_{\bk}{}^{'} \Psi^\dag_\bk  \mathcal{H}_B(\bk)  \Psi_{\bk},
\end{align}
where $\mathcal{H}_B(\bk)$ is a $2M\times2M$ matrix given by 
\begin{align}\label{BogMatrix}
&\mathcal{H}_B(\bk) = \begin{bmatrix}
	\mathcal{H}(\bk) -\mu_{\textrm{eff}} &  \Delta \\
	\Delta^* & \mathcal{H}^*(2\bk_c - \bk) -\mu_{\textrm{eff}},
\end{bmatrix}, \nonumber \\
& \Psi_\bk = [\delta c_{\bk 1}, \delta c_{\bk 2},..., \delta c_{\bk M}, \delta c^\dag_{2\bk_c -\bk 1},..., \delta c^\dag_{2\bk_c -\bk M}]^T, \nonumber \\
&[\Delta]_{\alpha\beta} = \delta_{\alpha,\beta} U n_0 \langle  \alpha | \phi_0 \rangle^2, \nonumber \\
&\mu_{\textrm{eff}} = (\epsilon_0 - \frac{Un_0}{M})\delta_{\alpha,\beta}.
\end{align}
The primed sum in Eq.~\eqref{ham} indicates that all the operators within the sum are for non-condensed states only, i.e. $\bk \neq \bk_c$ and $2\bk_c-\bk \neq \bk_c$.

To obtain the excitation energies from $H_B$, one needs to take care of the bosonic commutation rules by solving the eigenstates of $L(\bk) \equiv \sigma_z \mathcal{H}_B(\bk)$, where $\sigma_z$ is the Pauli matrix in the particle-hole space~\cite{castin:book}. One finds the excitation energies for each momentum $\bk$ to be $E_M(\bk) \geq ... E_2(\bk) \geq E_1(\bk) \geq 0 \geq -E_1(2\bk_c-\bk) \geq ... -E_M(2\bk_c-\bk)$. Here positive (negative) energies describe quasi-particle (-hole) excitations. The quasi-particle and -hole states are labelled as $|\psi^+_m (\bk) \rangle$ and $|\psi^-_m (\bk) \rangle$ such that 
\begin{align}
&L(\bk)|\psi^+_m(\bk) \rangle = E_m(\bk) |\psi^+_m(\bk) \rangle \\
&L(\bk)|\psi^-_m(\bk) \rangle = -E_m(2\bk_c - \bk) |\psi^-_m(\bk) \rangle. 
\end{align}
The chemical potential Eq.~\eqref{muexp} ensures that the lowest quasi-particle energy band is gapless at $\bk_c$, i.e. $E_1(\bk \rightarrow \bk_c) = 0$. This is the usual Goldstone mode that emerges because the condensate wavefunction acquires a complex phase and thus breaks the spontaneous gauge $U(1)$ symmetry~\cite{Fetter1971,Pitaevskii2003}. Therefore, Eq.~\eqref{muexp} can be thought as a generalization of the usual Hugenholtz-Pines relationship~\cite{Fetter1971} in case of a multiband system.

\section{Dependence of the speed of sound and excitation fraction on the orbital positions in multiband systems} \label{metricsection}

The speed of sound is defined as the slope of the lowest Bogoliubov excitation band around the condensate momentum $\bk_c$ 
\begin{align}
c_s = \lim_{\bq\rightarrow 0} \frac{E_1(\bk_c +\bq)}{|\bq|}
\label{speed_of_sound}
\end{align}
with $|\bq| \ll 0$. It was shown in Ref.~\cite{Julku2021a,Julku2021b} that in a flat band the low energy excitations of the condensate can be written as $E_1(\bk_c +\bq) = U n_0 \tilde{D}(\bq)/M$, where $\tilde{D}(\bq)$ is the condensate quantum distance. This quantity $\tilde{D}(\bq) \equiv \sqrt{1 - |\alpha(\bq)|^2}$ involves overlaps of the condensate state $|\phi_0\rangle\equiv |u_1(\bk_c)\rangle$ with neighboring Bloch states:
\begin{align}
    \alpha(\bq) \equiv M \sum_\alpha \langle u_1(\bk_c+\bq) | \alpha \rangle\langle \alpha| \phi_0 \rangle \langle \phi_0^*|\alpha \rangle \langle \alpha |u^*_1(\bk_c- \bq)\rangle  .
    \label{condensate_quantum_distance}
\end{align}
When $|u_1^*(\bk)\rangle = |u_1(\bk)\rangle$, the condensate quantum distance becomes identical to the usual Hilbert-Schmidt quantum distance and its infinitesimal limit $\tilde{D}(\bq) \rightarrow 4 \sum_{\mu\nu} q_\mu q_\nu g_{\mu\nu}(\bk_c)$ is proportional to the quantum metric
\begin{align}
g_{\mu\nu}(\bk_c)= \mathrm{Re}\left[ \langle \partial_\mu \phi_0 |(1-|\phi_0\rangle\langle \phi_0|)| \partial_\nu \phi_0\rangle\right] .
\label{metric}
\end{align}
In this way the speed of sound for a condensate in a flat band in Eq.~\eqref{speed_of_sound} is determined by the quantum metric
\begin{align}
c_s = \frac{2 U n_0}{M} \sqrt{g(\bk_c)} .
\label{speed_of_sound_metric}
\end{align}
This result can be contrasted to the fermionic counterpart where the flat band superconductivity is determined by the quantum metric integrated over the momemtum space~\cite{Peotta2015,Liang2017}. In contrast, here the speed of sound is determined by the quantum metric of the condensed Bloch state.

In a more general case $|u^*_1(\bk)\rangle \neq |u_1(\bk)\rangle$, the condensate quantum distance is expressed in terms of a generalized quantum metric 
\begin{align}
\tilde{g}_{\mu\nu}(\bk_c) = &\frac 1 2 \Big[ g_{\mu\nu}(\bk_c) + \mathrm{Re}\left(\langle \partial_\mu \phi_0|\phi_0\rangle\langle\phi_0|\partial_\nu \phi_0\rangle\right) +\label{generalizedmetric}\\
&2M\mathrm{Re}\left(\sum_\alpha \langle\partial_\mu \phi_0|\alpha\rangle\langle\alpha|\phi_0\rangle \langle\phi_0^*|\alpha\rangle\langle\alpha|\partial_\nu\phi_0^*\rangle \right)\Big]
\nonumber
\end{align}
and Eq.~\eqref{speed_of_sound_metric} holds with the replacement $g_{\mu\nu}(\bk_c) \rightarrow \tilde{g}_{\mu\nu}(\bk_c)$.
Similarly, the excitation fraction in the small interaction limit is also related to the condensate quantum distance $\tilde{D}(\bq)$ as
\begin{align}
\lim_{U\rightarrow 0} n_{ex}(\bk) = \lim_{U\rightarrow 0} \langle\delta c_\bk^\dagger \delta c_\bk\rangle = \frac{1-\tilde{D}(\bq)}{2\tilde{D}(\bq)},
\label{excitationfraction}
\end{align}
where $\bq = \bk - \bk_c$.

A potential problem when relating physical observable to the quantum metric and the quantum distance is the issue of orbital positioning in real space. In fact, the location of the atomic orbitals in multiband systems is crucial for the definition of the quantum metric~\cite{Huhtinen2022}, while the speed of sound and the excitation fraction do not depend on it. A physical explanation for the speed of sound is that the energy dispersion of the Bogoliubov Hamiltonian is obviously orbital-independent, therefore its slope cannot change when changing positions of lattice sites in real space. 
Here we show that the results stemming from the condensate distance $\tilde{D}(\bq)$ and the generalized quantum metric $\tilde{g}_{\mu\nu}$ are indeed orbital independent.
The orbital position can be taken explicitly into account when defining the Fourier transform as 
\begin{align}
c_{\bk\alpha} = \frac{1}{\sqrt{N}}\sum_i e^{-i \bk \cdot (\br_{i} + \br_\alpha)} c_{i\alpha},
\end{align}
where $\br_\alpha$ is the distance of the $\alpha$-th orbital from the position of the $i$-th unit cell $\br_{i}$. When $\br_\alpha=0$, the orbitals share the same real space position and the Hamiltonian is explicitly Bloch-periodic in reciprocal space. Let us consider for simplicity a 2 band-model, so that $M=2$, $|\alpha\rangle = \lbrace (1,0)^\transpose,(0,1)^\transpose\rbrace$. In this way the actual site position in real space is reflected in the Bloch states as an orbital-dependent phase factor, which can be written as 
\begin{align}
    |u(\bk)\rangle = \begin{pmatrix}u(\bk) e^{i \bk\cdot \br_\alpha}\\ 1\end{pmatrix}
\end{align} 
having omitted a trivial normalization factor that does not depend on $\br_\alpha$. 
Using this Bloch state for the calculation of the condensate quantum distance defined above, one can easily check that the condensate quantum distance $\tilde{D}(\bq)$ is orbital-independent, and so are the speed of sound obtained from the generalized metric in Eq.~\eqref{generalizedmetric} and the excitation fraction in Eq.~\eqref{excitationfraction} as obtained in \cite{Julku2021b}.
Notice that having the orbital-dependent phase factor means that $|u(\bk)\rangle \neq |u^*(\bk)\rangle$ so that the condensate distance is not equivalent to the usual Hilbert-Schmidt quantum distance. Therefore, taking the  infinitesimal limit of the condensate distance will not reduce to the quantum metric in Eq.~\eqref{metric}.

This result can be straightforwardly generalized to systems with more than two bands by noticing that each summand in Eq.~\eqref{condensate_quantum_distance} and Eq.~\eqref{generalizedmetric} is actually independent of the orbital position $\br_\alpha$.

To relate the speed of sound to the quantum metric, i.e.~to fulfill the condition $|u(\bk) \rangle = |u^*(\bk)\rangle$, one needs to find the orbital positions for which the hopping Hamiltonian $\mathcal{H}(\bk)$ is real as then $\mathcal{H}(\bk)|u^*(\bk)\rangle = \mathcal{H}^*(\bk)|u^*(\bk) \rangle = \Big( \mathcal{H}(\bk)|u(\bk) \rangle \Big)^* = \epsilon_\bk |u^*(\bk) \rangle$. For a non-degenerate band this implies $|u(\bk) \rangle \propto |u^*(\bk)\rangle$ such that $|u(\bk)\rangle$ can be chosen to be real with a trivial phase shift. The requirement for real-valued $\mathcal{H}(\bk)$ in turn means that for each hopping term from sublattice $\alpha$ to sublattice $\beta$ in spatial direction $\br$ with amplitude $t$, there must also exist a hopping term from sublattice $\alpha$ to $\beta$ to direction $-\br$ of amplitude $t^*$ as can be easily seen from Eq.~\eqref{hopping_element}. For real-valued hopping parameters, this condition requires the inversion symmetry to hold. For example, in case of kagome lattice it is easy to inspect that the geometry shown in Fig.~\ref{Fig:kagome}(a) fulfils this condition and thus with this choice of orbital positions the quantum metric can be related to the speed of sound, as was shown in Ref.~\cite{Julku2021a}. More generally in case of the complex-valued hopping parameters, the sufficient condition for $\mathcal{H}(\bk)$ to be real is the presence of the space-time inversion symmetry $C_2 T$.

\section{Superfluid weight of multiband bosons revisited} \label{sfwsection}

An important aspect regarding the flat band BEC is the phase coherence of the condensate. In two dimensional systems, this is related to the superfluid weight (or superfluid density) $D^s$ which determines the BKT phase transition temperature $T_{BKT}$ via $D^s(T_{BKT}) \propto T_{BKT}$. Superfluid weight of a flat band BEC at zero temperature was studied in Ref.~\cite{Julku2021b} in the case of the kagome lattice geometry with the linear response theory. Here we revisit the superfluid weight calculation, derive new terms to the superfluid weight, and point out possible problems related to the application of the Bogoliubov theory.

One can study the superfluid weight of the system by introducing a phase winding of momentum $\bq$ to the condensate wave function such that $\psi_0(\br_{i\alpha}) \rightarrow \psi_0(\br_{i\alpha}) e^{i \bq \cdot \br_{i\alpha}}$. The superfluid weight tensor $D^s_{\mu\nu}$ (where $\mu$ and $\nu$ denote the spatial indices) is then defined via the free energy $F$ of the system as
\begin{align}
D^s_{\mu\nu} = \frac{1}{N}\frac{d^2 F(\bq)}{dq_\mu dq_\nu} \Bigg|_{\bq=0,N_{\text{tot}}}.    
\end{align}
This approach is equivalent to introducing a vector potential $\bq$ to the kinetic Hamiltonian via Peierls substitution and then calculating the linear response to it and taking the $\bq=0$ limit.
The free energy reads $F(N_{\text{tot}},n_0,\bq) =  \Omega(\mu,n_0,\bq) +  N_{\text{\text{tot}}}\mu$, i.e. the thermodynamical variables for the grand-canonical potential are chemical potential $\mu$ and the condensate density $n_0$. Compared to the fermionic counterpart~\cite{Huhtinen2022}, the condensate density $n_0$ is analogous to the order parameter of the the Cooper pairs.

With the definition of the free energy  one can then in a straightforward manner obtain
\begin{align}
\frac{dF}{dq_\nu} &= \frac{d\Omega}{dq_\nu} + N_{\text{tot}}\frac{d\mu}{dq_\nu} \nonumber \\
&= \frac{\partial \Omega}{\partial q_\nu} + \frac{\partial \Omega}{ \partial n_0}\frac{dn_0}{dq_\nu} +    \frac{\partial \Omega}{ \partial \mu}\frac{d\mu}{dq_\nu} + N_{\text{tot}}\frac{d\mu}{dq_\nu} \nonumber \\
&=  \frac{\partial \Omega}{\partial q_\nu} + \frac{\partial \Omega}{ \partial n_0}\frac{dn_0}{dq_\nu}
\end{align}
where the last equality follows from $N_{\text{tot}} = - \frac{\partial \Omega}{\partial \mu}$. Now, taking the second derivative yields
the superfluid weight $D^s$ as follows
\begin{align}
\label{int_res}
&D^s_{\mu\nu} = \frac{1}{N} \frac{d^2 F}{dq_\mu dq_\nu} \Bigg|_{\bq=0} = D^s_{\text{old},\mu\nu} + D^s_{\text{corr},\mu\nu} + D^s_{\text{new},\mu\nu}, \nonumber \\
& [D^s_{\text{old}}]_{\mu\nu} = \frac{1}{N} \frac{\partial^2 \Omega}{\partial q_\mu q_\nu} \Bigg|_{\bq=0}, \nonumber \\
& [D^s_{\text{corr}}]_{\mu\nu} = \frac{1}{N} \Big[ \frac{\partial^2 \Omega}{\partial\mu \partial q_\nu}\frac{d\mu}{dq_\mu} +  \frac{\partial^2 \Omega}{\partial n_0 \partial q_\nu}\frac{dn_0}{dq_\mu} \Big]  \Bigg|_{N_{\text{tot}},\bq=0}, \nonumber \\
& [D^s_{\text{new}}]_{\mu\nu} = \frac{1}{N}\Big[\frac{d}{dq_\mu}\Big( \frac{\partial \Omega}{\partial n_0} \Big) \frac{d n_0}{dq_\nu} + \frac{\partial \Omega}{\partial n_0} \frac{d^2 n_0}{dq_\mu dq_\nu} \Big] \Bigg|_{N_{\text{tot}},\bq=0} .
\end{align}

Here the first term, i.e. $D^s_{\text{old},\mu\nu} \propto \frac{\partial^2 \Omega}{\partial q_\mu \partial q_\nu}$ has been extensively used in the literature to evaluate the mean-field superfluid density. In the Appendix~\ref{AppA1}, we show that $\frac{\partial^2 \Omega}{\partial q_\mu \partial q_\nu}$ is equivalent to the linear response formalism used in Ref.~\cite{Julku2021b} to study the superfluid weight of bosonic multiband systems. Similar conclusion was recently also reached in Ref.~\cite{Subasi2022}.

The second line of Eq.~\eqref{int_res} contains a correction  term $D^s_{\text{corr}}$ which was only recently pointed out to play an important role in the context  of fermionic superconductivity in multiband systems~\cite{Huhtinen2022}. In case of bosonic superfluidity, this correction can be also non-zero. Even though $D^s_{\text{corr}}$ includes the partial derivatives with respect to $\bq$, it turns out that these terms can be evaluated without the knowledge of the thermodynamic potential at finite $\bq$. This is shown in Appendix~\ref{AppB} where the expressions for $\frac{\partial^2 \Omega}{\partial q_\nu \partial \mu} \big|_{\bq=0}$ and $\frac{\partial^2 \Omega}{\partial q_\nu \partial n_0} \big|_{\bq=0}$ are derived. Furthermore, as discussed in Appendix~\ref{AppB}, the full derivatives $\frac{dn_0}{dq}$ and $\frac{d\mu}{dq}$ can be evaluated numerically in a straightforward manner.

Finally, the correction term in the third line of Eq.~\eqref{int_res} arises only in case of bosonic Bogoliubov theory and is absent in case of fermionic BCS superconductivity. This extra term comes from the fact that the partial derivative of the grand canonical potential with respect to $n_0$, i.e. $\frac{\partial \Omega}{ \partial n_0}$, does not necessarily vanish. This should be contrasted to the BCS theory where such partial derivatives with respect to the superconducting order parameter $\Delta$ are identically zero ($\partial \Omega /\partial \Delta = 0$) for any value of $\mathbf{q}$, yielding the usual BCS gap equation.
From the physical point of view, $\frac{\partial \Omega}{ \partial n_0}$ should vanish as it is the requirement for the minimization of the grand potential for a chosen $\mu$. Nevertheless, Bogoliubov theory can yield $\frac{\partial \Omega}{ \partial n_0} \neq 0$. This is a well-known problem~\cite{Yukalov2006,Griffin1996}, also called the Hohenberg-Martin dilemma, and it arises as the chemical potential is demanded to follow the Hughenholtz-Pines relationship to yield a gapless Goldstone mode at $\bk_c$. Thus, in order to employ Bogoliubov theory, one should always check how well the condition $\frac{\partial \Omega}{ \partial n_0} \sim 0$ is fulfilled. 

Let us now compute  $\frac{\partial \Omega}{ \partial n_0}$ in the case of the Bogoliubov theory. To this end, we note that $\frac{\partial \Omega}{\partial n_0} = \big\langle \frac{\partial H}{\partial n_0} \big\rangle$.
Applying this to the Bogoliubov Hamiltonian~\eqref{HPopov} yields
\begin{align}
\label{ddOdn0popov}
&\frac{1}{N} \frac{\partial \Omega}{\partial n_0} = \frac{1}{N}  \Bigg\langle \frac{\partial H}{\partial n_0} \Bigg\rangle = (\epsilon_0 - \mu)  + Un_0  \sum_\alpha |\langle \alpha | \phi_0 \rangle|^4 \nonumber \\
& + \frac{U}{2N} \sum_{\bk \alpha} \Big[4 | \langle \alpha | \phi_0 \rangle |^2\langle \delta c^\dag_{\bk \alpha} \delta c_{\bk \alpha} \rangle \nonumber \\
&+ \Big( \langle \alpha |  \phi_0  \rangle^2 \langle \delta c^\dag_{\bk \alpha} \delta c^\dag_{2\bk_c-\bk \alpha} \rangle + h.c. \Big) \Big].    
\end{align}
By using the expression~\eqref{muexp} for the chemical potential, we obtain
\begin{align}
\label{dOdno_popov}
&\frac{1}{N} \frac{\partial \Omega}{\partial n_0} =\frac{U}{2}\sum_\alpha \Big[ 4|\langle \alpha | \phi_0 \rangle|^2 n_{\text{ex,} \alpha} + \langle  \alpha |\phi_0  \rangle^2 \sigma^*_\alpha + h.c. \Big] \nonumber \\
& = \frac{2U n_{\text{ex}}}{M} + U\Re\Big[\sum_\alpha \langle  \alpha |\phi_0  \rangle^2 \sigma^*_\alpha\Big],  
\end{align}
where $n_{\text{ex}}$ ($n_{\text{ex,}\alpha}$) is the density of non-condensed particles (in sublattice $\alpha$), and $\sigma_\alpha = \frac{1}{N}\sum_\bk\langle \delta c^\dag_{\bk \alpha} \delta c^\dag_{2\bk_c-\bk \alpha} \rangle$. The last line in Eq.~\eqref{dOdno_popov} holds in case of the uniform BEC. Provided that the right hand side of Eq.~\eqref{dOdno_popov} is small compared to the other energy scales of the system such as hopping or interaction strength, one can approximate $\frac{1}{N} \frac{\partial \Omega}{\partial n_0} \approx 0$ which would fulfill the minimization of the thermodynamical energy requirement. Note that $\frac{1}{N}\frac{\partial^2 \Omega}{\partial n_0^2} = U \sum_\alpha |\langle \alpha | \phi_0 \rangle|^4 >0$ so the state would be thermodynamically stable.

The fact that $D^s_{\text{new}}$  arises from the flaw of the Bogoliubov theory highlights the necessity to go beyond Bogoliubov theory when the quantum fluctuation terms start to be significant. One way to circumvent the Hohenberg-Martin dilemma is to introduce an additional Lagrange multiplier for the condensed particles~\cite{Yukalov2006}. Effectively, this means to use different chemical potentials for the BEC and the non-condensed particles. Another possibility is of course to use more advanced methods such as exact diagonalization~\cite{Huber2010}, quantum Gutzwiller theory~\cite{Caleffi2020} or quantum Monte Carlo methods~\cite{Sadoune2022}. We leave these aspects for future studies.

\begin{figure}
  \centering
    \includegraphics[width=1.0\columnwidth]{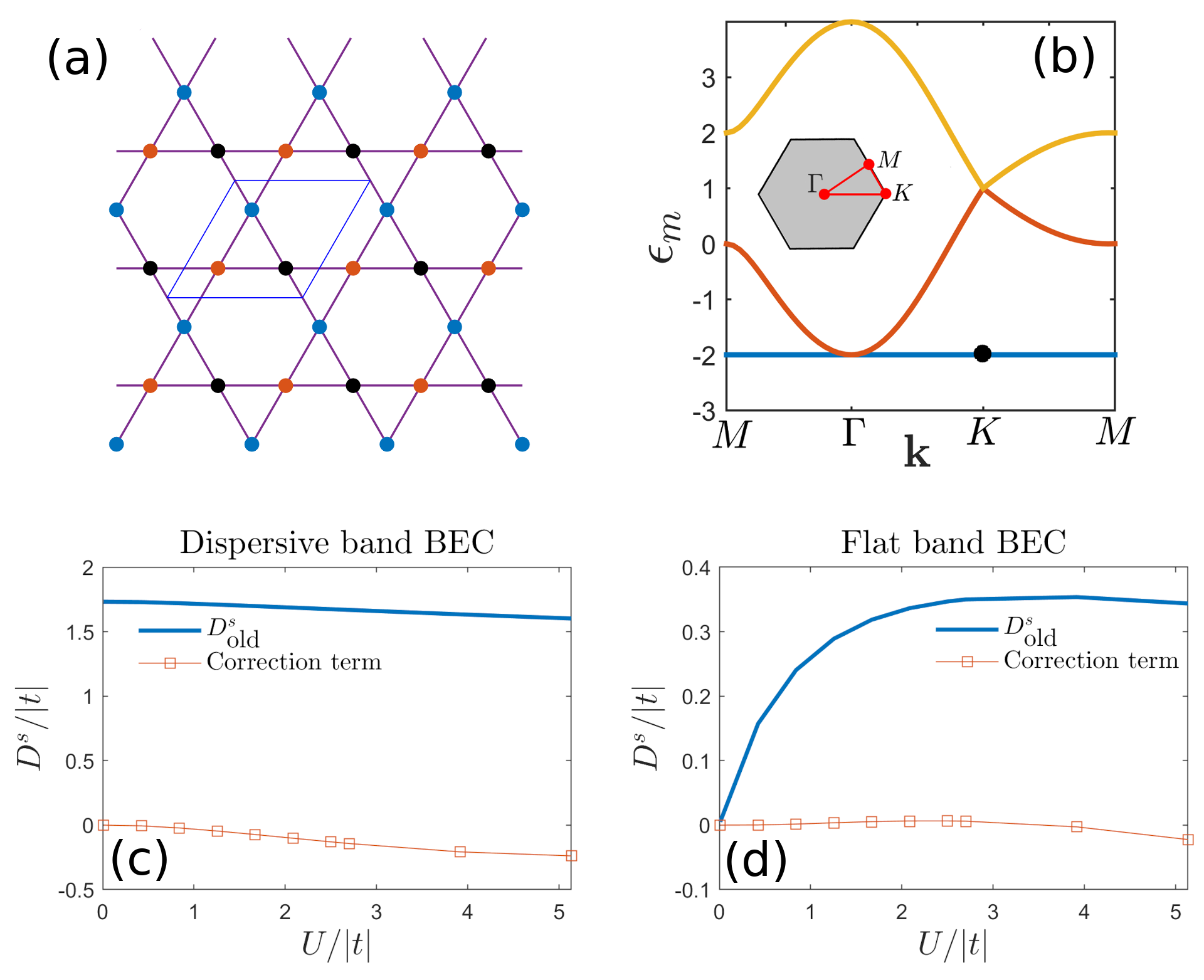}
    \caption{(a) Kagome lattice geometry. Black, red and blue circles depict the sites of sublattices $\alpha =$ 1, 2 and 3, respectively. Purple lines present the nearest-neighbour (NN) hopping strength of $t$. The unit cell is shown as a blue parallelogram. (b) One-particle energy band structure for $t=1$ as a function of momentum along the path connecting the high-symmetry points $\Gamma$, $K$ and $M$. The lowest band is flat for positive $t$ and the flat band BEC can take place at $\bk = [4\pi/3,0]$ denoted as a black dot. Inset shows the hexagonal Brillouin zone as well as the high-symmetry points. (c)-(d) Superfluid weight terms $D^s_{\text{old}}$ and $D^s_{\text{new}}$ for the dispersive band BEC ($t<0$) and  flat band BEC ($t>0$), respectively.  Total density is fixed to $n_{\text{tot}} =3$. We see that the unphysical term $D^s_{\text{new}}$ is small compared to the total superfluid weight at the weak-coupling regime.}
   \label{Fig:kagome}
\end{figure}

As an example, we study now the superfluid weight in the kagome lattice geometry [see Fig.~\ref{Fig:kagome}(a)] that was also considered in Ref.~\cite{Julku2021b}. This system is characterized by three sublattices and nearest-neighbour (NN) hopping parameter $t$. If one has $t>0$, the lowest Bloch bands is flat as shown in Fig.~\ref{Fig:kagome}(b). By inverting the sign of $t$, the band structure is flipped and one of the dispersive bands becomes the lowest Bloch band. In this way one can study both the  dispersive band and flat band BECs. For the dispersive band BEC ($t<0$), the condensate takes place at zero momentum, i.e., $\bk_c = 0$ with the corresponding Bloch state being $|\phi_0 \rangle = [1,1,1]$. In case of the flat band BEC ($t>0$), the condensate can take place at the corners of the hexagonal Brilliouin zone, for example at the $K$-point such that $\bk_c = [4\pi/3,0]$ with $|\phi_0 \rangle = [-1,-1,1]$~\cite{Huber2010,you:2012,Julku2021a,Julku2021b}. Thus, the uniform BEC condition, i.e. $|\langle \alpha | \phi_0 \rangle|^2 = 1/M$ is fulfilled in both cases. 

In Figs.~\ref{Fig:kagome}(c)-(d) we show the superfluid weight both in case of dispersive band and flat band BEC in kagome lattice, respectively, as a function of interaction $U$ at zero temperature. It turns out that for the kagome lattice BEC we have $\frac{dn_0}{dq} = \frac{d\mu}{dq} = 0$ such that  $D^s_{\text{corr}} =0$. We therefore plot $D^s_{\text{old}}$ and and correction term $D^s_{\text{new}}$. We see that the unphysical term $D^s_{\text{new}}$ is much smaller compared to $D^s_{\text{old}}$ in both cases in the weak-coupling regime, implying that the flaw of the Bogoliubov theory, i.e. the violation of the condition $\frac{\partial \Omega}{ \partial n_0} = 0$, should not play a significant role in the regime of weak interactions. At larger interactions $D^s_{\text{new}}$ becomes more significant, signalling the breakdown of the Bogoliubov theory.

From Figs.~\ref{Fig:kagome}(c)-(d) we can see, as pointed out also in Ref.~\cite{Julku2021b}, that the origin of the superfluid weight is very different in case of dispersive and flat band condensates. Namely, in case of the dispersive band $D^s$ is maximized at $U=0$. This can be understood by the fact that in the dispersive band it is the condensed particles that carry the superfluid flow. Increasing interactions yield larger quantum depletion and thus smaller condensate density. In contrast, in the case of a flat band BEC, $D^s$ vanishes at $U=0$ and becomes only finite for finite interactions. Thus, quantum fluctuations actually trigger finite flat band superfluidity instead of suppressing it. Physically this can be understood by the fact that non-interacting bosons in a flat band are strongly localized and thus their group velocity is zero. Consequently, flat band superfluid weight vanishes at the non-interacting limit. From Fig.~\ref{Fig:kagome}(d) one can also see that the superfluid weight of the flat band BEC is actually a non-monotonic function of $U$, reaching its maximum around $U\sim 3t$ and decreasing for larger interaction strengths, in a similar manner as in case of the dispersive band BEC. Physically, this makes sense as at $U\rightarrow \infty$ the bosons become essentially hard-core bosons and thus increasingly more localized as a function of $U$. Consequently, the system loses its long-range coherence and becomes an insulator.

%%%%%%%%%%%%%%%%%%%%%%%%%%%%%%%%%%%%%%%%%%%%%%%%%%%%%%%%%%%%%%%%%%%%%%%%%%%
%%%%%%%%%%%%%%%%%%%%%%%%%%%%%%%%%%%%%%%%%%%%%%%%%%%%%%%%%%%%%%%%%%%%%%%%%%%

%%%%%%%%%%%%%%%%%%%%%%%%%%%%%%%%%%%%%%%%%%%%%%%%%%%%%%%%%%%%%%%%%%%%%%%%%%%

%%%%%%%%%%%%%%%%%%%%%%%%%%%%%%%%%%%%%%%%%%%%%%%%%%%%%%%%%%%%%%%%%%%%%%%%%%%

\section{Second order correction to the speed of sound}\label{Sec_cs}

As discussed in Sec.~\ref{metricsection}, the Goldstone excitation energy of a uniform flat band BEC can be approximated as $E_1(\bk_c + \bq) \sim \frac{Un_0}{M}\tilde{D}(\bq)$ in the weak-coupling limit. This result was derived in Refs.~\cite{Julku2021a,Julku2021b} by expressing the Bogoliubov Hamiltonian $\mathcal{H}_B(\bk)$ (or equivalently $L(\bk)= \sigma_z \mathcal{H}_B(\bk)$) in the Bloch basis and then discarding other than the flat band degrees of freedom. This yields the $2\times2$ projected matrix
\begin{align}
\label{lk2}
L_p(\bk) =  \begin{bmatrix}  \frac{Un_0}{M} & \frac{Un_0}{M}\alpha(\bq) \\
- \frac{Un_0}{M}\alpha^*(\bq)  & -\frac{Un_0}{M}
\end{bmatrix}. 
\end{align}
By diagonalizing $L_p$ at the small $\bq$ limit, one finds the dispersion $E_1(\bk_c + \bq) = \frac{Un_0}{M}\tilde{D}(\bq)$ for the gapless Goldstone mode. However, by taking into account the coupling between flat band and other higher energy Bloch bands up to the second order, one can derive a correction term for $E_1(\bk)$ that holds at a larger interaction regime. If one denotes $E^{(0)}_\bq \equiv \frac{Un_0}{M}\tilde{D}(\bq)$, and applies the standard 2nd order perturbation theory, one finds for the Goldstone mode in the $\bq \rightarrow 0$ limit (for details, see Appendix~\ref{AppB})
\begin{align}
&E_1(\bk_c + \bq) \approx E^{(0)}_\bq + E^{(2)}_\bq   \quad \text{with} \nonumber \\
&E^{(2)}_\bq = -\sum_{j=2}^M \frac{\tilde{U}|\langle u_1(\bk_c-\bq) |\Delta| u^*_j(\bk_c+\bq) \rangle|^2}{2 E^{(0)}_\bq[ \tilde{\epsilon}_{j}(\bk_c + \bq)  + \tilde{U} ]}   \nonumber \\
 & -\sum_{j=2}^M \frac{\tilde{U}|\langle u_1(\bk_c+\bq) |\Delta| u^*_j(\bk_c-\bq) \rangle|^2}{2 E^{(0)}_\bq[ \tilde{\epsilon}_{j}(\bk_c - \bq) + \tilde{U} ]}, \\ \nonumber
& \tilde{U} \equiv Un_0/M \nonumber \\
& \tilde{\epsilon}_j(\bk) \equiv \epsilon_j(\bk) - \epsilon_1(\bk_c) .
\end{align}
By assuming that the energy gap from the flat Bloch band to all the higher bands is sufficiently large, we can approximate $\tilde{\epsilon}_j(\bk_c) \approx \Delta_g$, where $\Delta_g$ is the average gap between the flat band and higher bands at $\bk_c$, i.e. $\Delta_g = \frac{1}{M-1}\sum_{j=2}^M \tilde{\epsilon}_j(\bk_c)$. With this approximation, we write 
\begin{align}
E^{(2)}_\bq &\approx - \frac{\tilde{U}\tilde{U}^2( 1 - |\alpha(\bq)|^2) }{E^{(0)}_\bq [\tilde{U} + \Delta_g]} \nonumber \\
&  = - \tilde{U}\frac{E^{(0)}_\bq}{\tilde{U} + \Delta_g}
\end{align}
where  in the first line we have exploited the identity $1 = \sum_{j=1}^M | u^*_j(\bk_c - \bq)\rangle \langle u^*_j(\bk_c - \bq) |$ and in the second line used $E^{(0)}_\bq = \tilde{U}\sqrt{1 - |\alpha(\bq)|^2}$. Importantly, we see that the correction term is proportional to $E^{(0)}_\bq$, i.e. the speed of sound up to the second order in interaction reads
\begin{align}
&c_s  = c_s^{(0)}(1 - \frac{1}{1 + \frac{\Delta_g}{\tilde{U}}}), \quad \text{with}\label{cs2} \\
&c_s^{(0)} = \frac{2 U n_0}{M} \sqrt{\tilde{g}(\bk_c)} \label{cs0},
\end{align}
where $c_s^{(0)}$ is the result derived in Ref.~\cite{Julku2021b}. In other words, the second order result of the speed of sound, $c^{(2)}_s$, is still determined by the generalized quantum metric $\tilde{g}$. Furthermore, from the expression of $c_s$ we note that the correction term becomes smaller when the ratio $\Delta_g/\tilde{U}$ increases, as expected.

\begin{figure}
  \centering
    \includegraphics[width=1.0\columnwidth]{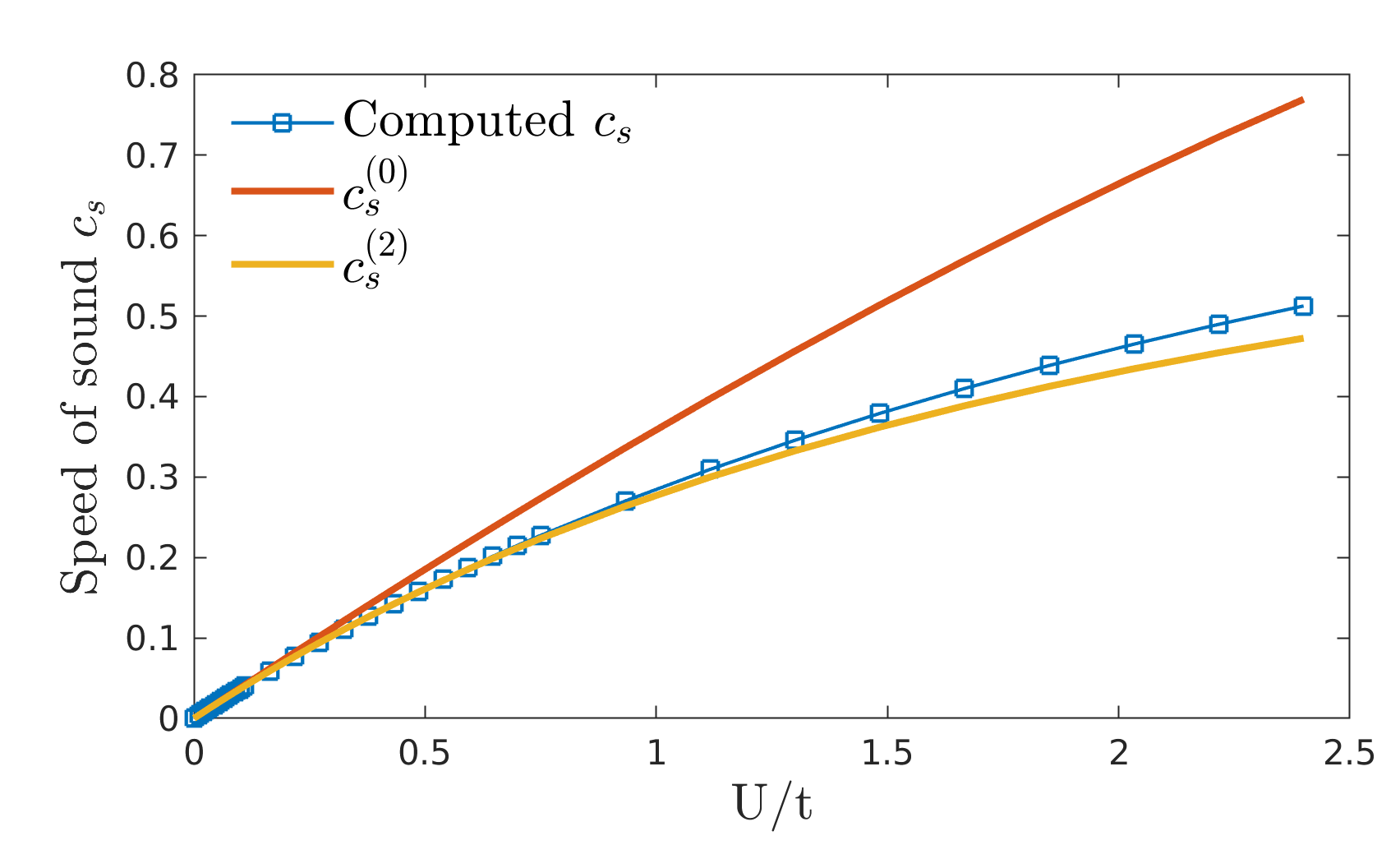}
    \caption{Speed of sound for the flat band BEC in the kagome lattice as a function of interaction $U$. The blue squares denote the full Bogoliubov result, whereas $c^{(2)}_s$ and $c^{(0)}_s$ denote the predictions of Eqs.~\eqref{cs2} and \eqref{cs0}, respectively. The total density is taken to be $n_{\text{tot}} = 3$.}
   \label{Fig:cs}
\end{figure}

To see how well the 2nd order correction relates to the numerical result, in Fig.~\ref{Fig:cs} we show $c_s$ of a flat band BEC in the case of the kagome  lattice as a function of interaction. By comparing the full Bogoliubov result to the estimates given by Eqs.~\eqref{cs2} and \eqref{cs0}, we see that the second order result $c^{(2)}_2$ works well even with larger interaction strengths.

%%%%%%%%%%%%%%%%%%%%%%%%%%%%%%%%%%%%%%%%%%%%%%%%%%%%%%%%%%%%%%%%%%%%%%%%%%%
%%%%%%%%%%%%%%%%%%%%%%%%%%%%%%%%%%%%%%%%%%%%%%%%%%%%%%%%%%%%%%%%%%%%%%%%%%%

\section{Conclusions} \label{conclusions}
The stability of a BEC in a flat band has been shown to be determined by quantum geometry~\cite{Julku2021a,Julku2021b}. Here we have clarified that the speed of sound of the BEC and the excitation fraction (quantum depletion) are, in general, governed by generalized versions of the quantum metric and the quantum distance, which both are independent of orbital positions. Only in lattices of sufficiently high symmetry there is a connection from the BEC properties to the usual quantum metric and the Hilbert-Smith quantum distance. This relates to the recent findings regarding fermionic superfluidity and superconductivity in a flat band where the superfluid weight is governed by the minimal quantum metric, which coincides with the usual quantum metric when the orbital positions are related by a symmetry operation~\cite{Huhtinen2022,Herzog-Arbeitman2022}. We furthermore presented a second order correction to the analytical formula that connects the speed of sound and quantum geometric quantities, showing good agreement with numerical results.

When calculating the superfluid weight in fermionic superfluid systems, it was shown that it is essential to consider the dependence on the Cooper pair momentum $\mathbf{q}$ (equivalently, the vector potential) for all self-consistently calculated quantities~\cite{Huhtinen2022,Herzog-Arbeitman2022}. This motivated us to revisit the superfluid weight calculation in the BEC case by taking into account the full dependence of the condensate order parameter $n_0$ and the chemical potential $\mu$ on $\mathbf{q}$. As in the case of fermions, also the bosonic superfluid features a correction term [$D^s_{\text{corr}}$ in Eq.~\eqref{int_res}] stemming from the $\bq$-dependence of $\mu$ and $n_0$. Moreover, we showed that in the framework of the Bogoliubov theory, there is also an additional term, $D^s_{\text{new}}$, that is unique to the bosonic case. This term arises due to the known flaw of Bogoliubov theory, namely that the grand potential is not a minimum of the condensate density ($\partial \Omega/\partial n_0 \neq 0$). While these new additional terms of superfluid weight derived in our present work do not change the results of the previous work~\cite{Julku2021b} significantly, our results suggest that further work is needed to derive the superfluid weight within an approach where the flaws of Bogoliubov theory have been fixed. Exact numerical approaches might be a helpful guide in developing such theory.

\section*{Acknowledgements}
We thank Kukka-Emilia Huhtinen and Georg Bruun for useful discussions. P. T. and G.S.~acknowledge support by the Academy of Finland under project numbers 307419, 327293, 349313. This work has been supported by the Danish National Research Foundation through the Center of Excellence “CCQ” (Grant agreement no.: DNRF156). G.S. has received funding from the European Union's Horizon 2020 research and innovation program under the Marie Sk\l{}odowska-Curie grant agreement N. 101025211 (TEBLA).

\begin{widetext}
\appendix
\section{Details on the Bogoliubov approximation}\label{AppA}

Bogoliubov theory can be conveniently derived  in the Green's function formalism. In Ref.~\cite{Julku2021b} this derivation was provided in case of multiband lattice systems. Here we show an alternative way to derive the Bogoliubov theory without the Green's functions, to make it easier to obtain the superfluid weight  in the phase winding formalism.

Our starting point is Eq.~\eqref{spoint} which we re-express here in the momentum space as
\begin{align}
\label{ckop}
&c_{\bk \alpha} = \frac{1}{\sqrt{N}}\sum_{\br_{i\alpha}}e^{-i\bk \cdot \br_{i\alpha}}c_{i\alpha} \nonumber \\
&= \sqrt{Nn_0} \langle \alpha | \phi_0 \rangle \delta_{\bk,\bk_c} +  \frac{1}{\sqrt{N}}\sum_{\br_{i\alpha}}e^{-i\bk \cdot \br_{i\alpha}}\delta c_{i\alpha} \nonumber \\
&\equiv \sqrt{Nn_0}\langle \alpha | \phi_0 \rangle \delta_{\bk,\bk_c} + \delta c_{\bk\alpha},
\end{align}
and correspondingly in the Bloch basis we have
\begin{align}
&\gamma_{\bk n} = \sum_\alpha [U^\dag(\bk)]_{n\alpha}c_{\bk\alpha} = \nonumber \\
&\sqrt{Nn_0}\delta_{\bk,\bk_c}\delta_{n,1} + \sum_\alpha [U^\dag(\bk)]_{n\alpha}\delta c_{\bk\alpha} \nonumber \\
&\equiv \sqrt{Nn_0}\delta_{\bk,\bk_c}\delta_{n,1} + \delta \gamma_{\bk n}.   
\end{align}

We now use these expressions in Eq.~\eqref{BHH} and discard third and fourth order fluctuation terms. In order to have the condensation as a stable ground state, we first demand that the terms linear to the BEC fluctuation operators, i.e. $\delta \gamma_{\bk_c,1}$, vanish, i.e.
\begin{align}
& 0 = \sqrt{N n_0}[\epsilon_1(\bk_c) -\mu](\delta \gamma_{\bk_c 1} + \delta \gamma^\dag_{\bk_c1}) \nonumber \\
&+ \frac{U}{2N}\sum_\alpha |\langle \alpha | \phi_0 \rangle|^4 \Big[2(Nn_0)^{3/2}\delta \gamma_{\bk_c1} + 2(Nn_0)^{3/2}\delta \gamma^\dag_{\bk_c1}\Big]  \nonumber \\
&\Leftrightarrow 0 = \sqrt{N n_0}(\epsilon_1(\bk_c) -\mu) + U \sqrt{N}n_0^{3/2} \sum_\alpha |\langle \alpha | \phi_0 \rangle|^4.
\end{align}
This yields 
\begin{equation}
\mu = \epsilon_0 + Un_0 \sum_\alpha |\langle \alpha | \phi_0 \rangle|^4, \label{eq:mu_in_App}
\end{equation}
i.e. Eq.~\eqref{muexp}. We treat the interaction Hamiltonian within the Bogoliubov approximation by taking into account only the scattering processes where two or four condensation lines enter/leave the interaction
vertex. In other words, we keep only the interaction terms that are quadratic order in the fluctuations $\delta c_{\bk\alpha}$ and $\delta c^\dag_{i\alpha}$. The full Hamiltonian is then
\begin{align}
H = H_c + H_{c-f} + H_f    
\end{align}
where $H_c$ describes the condensate, $H_{c-f}$ is the coupling between the condensate and fluctuations and $H_f$ arises solely from the fluctuations. The condensate term reads
%\begin{widetext}
\begin{align}
\label{origHc}
&H_c = n_0\sum_{i\alpha j\beta} ( t_{i\alpha,j\beta} -\mu\delta_{i\alpha,j\beta} )e^{-i\bk_c\cdot (\br_{i\alpha} - \br_{j\beta})}\langle \phi_0 | \alpha \rangle \langle \beta | \phi_0 \rangle  + \frac{Un_0^2 N}{2} \sum_\alpha|\langle \alpha | \phi_0 \rangle|^4 \nonumber \\
& =(\epsilon_1(\bk_c) - \mu)N_0 + 
\frac{Un_0^2 N}{2} \sum_\alpha|\langle \alpha | \phi_0 \rangle|^4 = -\frac{U n_0 N_0}{2} \sum_\alpha|\langle \alpha | \phi_0 \rangle|^4,
\end{align}
%\end{widetext}
where $N_0 = n_0 N$ is the number of condensed particles and in the second line we have exploited the expression~\eqref{eq:mu_in_App} (\eqref{muexp} in the main text) for the chemical potential. The condensate-fluctuation term is zero but for later use we write it explicitly as
\begin{align}
H_{c-f} &= \sqrt{n_0}\sum_{i\alpha j\beta} \big [ (t_{i\alpha,j\beta} -\mu\delta_{i\alpha j\beta} ) e^{-i\bk_c \cdot \br_{i\alpha}} \langle \phi_0 | \alpha \rangle \delta c_{j\beta} + h.c. \big] \nonumber \\
&+ Un_0^{3/2} \sum_{i\alpha} |\langle \alpha | \phi_0 \rangle|^2 \big ( e^{-i\bk_c \cdot \br_{i\alpha}} \langle \phi_0 | \alpha \rangle \delta c_{i\alpha} + h.c. \big )  =0.
\end{align}
Mathematically the equality $H_{c-f}=0$ can be understood by first realizing that the kinetic term $H_0$ does not have terms linear in $\delta \gamma_{\bk n}$ when $\bk \neq \bk_c$ or $n\neq 1$ at $\bk = \bk_c$. Furthermore, the momentum conservation  ensures that linear terms for $\bk \neq \bk_c$ cannot exist in the interaction term either as setting three momenta in the interaction vertex to $\bk_c$ would necessarily force also the remaining momentum to be $\bk_c$. This still leaves possibility for having linear terms in $\delta \gamma_{\bk_c,n>1}$ in the interaction term. These terms vanish for uniform condensates, i.e. when $|\langle \alpha | \phi_0 \rangle|^2 = 1/M$. In the following we assume this and thus take  $H_{c-f}=0$.

Finally, the fluctuation Hamiltonian is
%\begin{widetext}
\begin{align}
\label{origHf}
H_f = \sum_{i\alpha j\beta} \big [ (t_{i\alpha,j\beta} -\mu\delta_{i\alpha j\beta} ) \delta c^\dag_{i\alpha}\delta c_{j\beta} + \frac{U}{2}\sum_{i\alpha}\big[ e^{2i\bk_c \cdot \br_{i\alpha}} n_0 \langle \alpha | \phi_0 \rangle^2 \delta c^\dag_{i\alpha} \delta c^\dag_{i \alpha} + h.c + 4Un_0 | \langle \alpha | \phi_0 \rangle |^2 \delta c^\dag_{i\alpha} \delta c_{i\alpha} \big]
\end{align}
%\end{widetext}

Even though the condensate term is a simple constant and the condensate-fluctuation term vanishes, it is still important to know their form as we will calculate the superfluid weight by introducing the phase winding to the condensate wave function. This will then affect the kinetic terms of $H_c$ and $H_{c-f}$ (actually it also affects the interaction of $H_{c-f}$ but this does not have a contribution in the superfluid weight calculation).

\subsection{Superfluid weight}\label{AppA1}
In this section we derive the superfluid contribution $D^s_{\text{old}} = \frac{\partial^2 \Omega}{\partial q_\mu \partial q_\nu}\Bigg|_{\bq=0}$ and show that this is equivalent to the linear response result derived in Ref.~\cite{Julku2021b}. To this end, we introduce the phase winding as $\psi_0(\br_{i\alpha}) \rightarrow \psi_0(\br_{i\alpha}) e^{i \bq \cdot \br_{i\alpha}}$ and, for convenience, we further apply the basis transformation $ \delta c_{i\alpha} \rightarrow e^{i \bq \cdot \br_{i\alpha}} \delta c_{i\alpha}$. We then find for the condensate term at small $\bq$ the following:
%\begin{widetext}
\begin{align}
\label{condterm}
&H_c(\bq) = n_0\sum_{i\alpha j\beta}( t_{i\alpha j\beta} - \mu \delta_{i\alpha,j\beta} )e^{-i\bk_c\cdot (\br_{i\alpha} - \br_{j\beta})} e^{-i\bq\cdot (\br_{i\alpha} - \br_{j\beta})} \langle \phi_0 | \alpha \rangle \langle \beta | \phi_0 \rangle +\frac{U n_0 N_0}{2} \sum_\alpha|\langle \alpha | \phi_0 \rangle|^4 \nonumber \\
&\approx n_0 \sum_{i\alpha j\beta} ( t_{i\alpha j\beta} - \mu \delta_{i\alpha,j\beta} )e^{-i\bk_c\cdot (\br_{i\alpha} - \br_{j\beta})} \big[ 1 - i \bq \cdot \br_{i\alpha j\beta} -\frac{1}{2}(\bq \cdot \br_{i\alpha j\beta})^2 \big]  \langle \phi_0 | \alpha \rangle \langle \beta | \phi_0 \rangle +\frac{U n_0 N_0}{2} \sum_\alpha|\langle \alpha | \phi_0 \rangle|^4 \nonumber \\
&= \sum_\mu q_\mu \Big[ n_0 N \langle \phi_0 | \partial_\mu \mathcal{H}(\bk_c) | \phi_0 \rangle + \frac{N n_0}{2} \sum_\nu \langle \phi_0 | \partial_\mu \partial_\nu \mathcal{H}(\bk_c) | \phi_0 \rangle q_\nu \Big] + H_c \nonumber \\
& = \sum_\mu q_\mu \frac{N n_0}{2} \sum_\nu \langle \phi_0 | \partial_\mu \partial_\nu \mathcal{H}(\bk_c) | \phi_0 \rangle q_\nu  + H_c,    
\end{align}
%\end{widetext}
where we have denoted $\partial_\mu \equiv \frac{\partial}{\partial k_\mu}$ and $H_c$ is the original condensate Hamiltonian Eq.~\eqref{origHc}. In the second line the phase winding term is expanded up to second order in $\bq$ and the last line follows from the vanishing current in the BEC ground state when $\mathbf{q}=0$, i.e.~$\langle \phi_0 | \partial_\mu \mathcal{H}(\bk_c) | \phi_0 \rangle =0$. In the same way, we have for the kinetic part of the condensate-fluctuation term 
%\begin{widetext}
\begin{align}
\label{condflucterm}
H_{c-f,kin}(\bq) &= \sqrt{n_0}\sum_{i\alpha j\beta} ( t_{i\alpha j\beta} - \mu \delta_{i\alpha,j\beta} ) \langle \phi_0 | \alpha \rangle e^{-i\bk_c \cdot \br_{i\alpha}} e^{-i\bq \cdot \br_{i\alpha}} e^{i\bq \cdot \br_{j\beta}} \delta c_{j\beta} \nonumber \\
&\approx \sqrt{n_0}\sum_{i\alpha j\beta} ( t_{i\alpha j\beta} - \mu \delta_{i\alpha,j\beta} ) \langle \phi_0 | \alpha \rangle e^{-i\bk_c \cdot \br_{i\alpha}}  \big[ 1 - i \bq \cdot \br_{i\alpha j\beta} -\frac{1}{2}(\bq \cdot \br_{i\alpha j\beta})^2 \big] \delta c_{j\beta} \nonumber \\
&= H_{c-f,kin} +  \sqrt{N_0}\sum_\mu q_\mu \Big[ \sum_{\alpha\beta}( \langle \phi_0 | \alpha \rangle \partial_\mu \mathcal{H}_{\alpha\beta}(\bk_c) \delta c_{\bk_c \beta} + h.c. )  +\frac{1}{2} \sum_\nu q_\nu  \sum_{\alpha\beta}( \langle \phi_0 | \alpha \rangle \partial_\mu \partial_\nu \mathcal{H}_{\alpha\beta}(\bk_c) \delta c_{\bk_c\beta} + h.c.)  \Big] \nonumber \\
& = \sqrt{N_0}\sum_\mu q_\mu \Big[ \sum_{\alpha\beta}( \langle \phi_0 | \alpha \rangle \partial_\mu \mathcal{H}_{\alpha\beta}(\bk_c) \delta c_{\bk_c \beta} + h.c. )  +\frac{1}{2} \sum_\nu q_\nu  \sum_{\alpha\beta}( \langle \phi_0 | \alpha \rangle \partial_\mu \partial_\nu \mathcal{H}_{\alpha\beta}(\bk_c) \delta c_{\bk_c\beta} + h.c.),
\end{align}
%\end{widetext}
where in the last line we have used $H_{c-f,kin} = 0$. Similarly, the kinetic part of the fluctuation term becomes in the new basis up to the second order in $\bq$
%\begin{widetext}
\begin{align}
\label{flucterm}
&H_{f,kin}(\bq) = \sum_{i\alpha j\beta}( t_{i\alpha j\beta} - \mu \delta_{i\alpha,j\beta} ) e^{-i \bq \cdot (\br_{i\alpha} - \br_{j\beta})} \delta c^\dag_{i\alpha} \delta c_{j\beta}  \approx \sum_{i\alpha j\beta}( t_{i\alpha j\beta} - \mu \delta_{i\alpha,j\beta} ) \big[ 1 - i \bq \cdot \br_{i\alpha j\beta} -\frac{1}{2}(\bq \cdot \br_{i\alpha j\beta})^2 \big] \delta c^\dag_{i\alpha} \delta c_{j\beta} \nonumber \\
&= H_{f,kin}  + \sum_\mu q_\mu \Big[ \sum_\bk \delta c^\dag_\bk \partial_\mu \mathcal{H}(\bk) \delta c_\bk 
+ \frac{1}{2}\sum_{\bk\nu} \delta c^\dag_\bk \partial_\mu \partial_\nu \mathcal{H}(\bk) \delta c_\bk  q_\nu \Big], 
\end{align}
%\end{widetext}
where $[\delta c_\bk]_\alpha = \delta c_{\bk \alpha}$ is a vector containing the fluctuation operators. From Eqs.~\eqref{condterm},~\eqref{condflucterm} and~\eqref{flucterm} we can easily see that the total Hamiltonian can be written as $H(\bq) = H + H'(\bq)$. 

We expand now the grand canonical potential $\Omega(\bq)$ as a perturbation series with respect to $\bq$ and keep the terms up to second order in $\bq$. To this end, we need to first evaluate the partition function $Z(\bq)$ which can be written as
\begin{align}
Z(\bq) &= \text{Tr} [ e^{-\beta H(\bq)}] =  \text{Tr} [ e^{-\beta H} (1 + \tilde{U}(\beta,0))] \nonumber \\
&= \text{Tr} [e^{-\beta H}] + \text{Tr} [e^{-\beta H} \tilde{U}(\beta,0)] \nonumber \\
& = Z(0) + \text{Tr} [e^{-\beta H} \tilde{U}(\beta,0)],
\end{align}
with 
\begin{align}
\label{lce}
&\tilde{U}(\beta,0) = \nonumber \\
&\sum_{n=1}^{\infty} \frac{(-1)^n}{n!}  \int_0^\beta d\tau_1 \cdots \int_0^\beta d\tau_n  T_\tau H'(\bq,\tau_1) \cdots H'(\bq,\tau_n) , 
\end{align}
%where the expectation value $\langle \cdots \rangle$ is taken with respect to the $\bq=0$ state. 
We expand $\tilde{U}(\beta,0)$ up to the second order in $\bq$ which gives 
%\begin{widetext}
\begin{align}
\label{partZ}
Z(\bq) &= Z(0) + Z(0) \Big[ -\int_0^\beta d\tau_1 \langle T_\tau H'(\bq,\tau_1) \rangle + \frac{1}{2} \int_0^\beta d\tau_1 \int_0^\beta d\tau_2 \langle T_\tau H'(\bq,\tau_1) H'(\bq,\tau_2) \rangle \Big] + \mathcal{O}(q^3) \nonumber \\
& \equiv Z(0)[1 + z_1(\bq) + z_2(\bq)] + \mathcal{O}(q^3) ,    
\end{align}
%\end{widetext}
where the expectation value $\langle \cdots \rangle$ is taken with respect to the $\bq=0$ state. After solving the partition function $Z(\bq)$ from Eq.~\eqref{partZ}, one can compute the grand canonical potential as
\begin{align}
\label{gcp}
&\Omega(\bq) = -\frac{1}{\beta}\log Z(\bq) \nonumber \\
&\approx -\frac{1}{\beta}\log Z(0)  - \frac{1}{\beta}\log (1 + z_1(\bq) + z_2(\bq) ) \nonumber \\
&= \Omega(\bq=0) - \frac{1}{\beta}\log (1 + z_1(\bq) + z_2(\bq) ).
\end{align}
so that 
\begin{align}
\label{ddO}
\frac{\partial^2 \Omega(\bq)}{\partial q_\mu \partial q_\nu}\Bigg|_{\bq=0} =& \frac{1}{\beta} \frac{\partial [z_1(\bq) + z_2(\bq)]}{\partial q_\mu} \frac{\partial [z_1(\bq) + z_2(\bq)]}{\partial q_\nu}\Bigg|_{\bq=0} \nonumber \\
&- \frac{1}{\beta} \frac{\partial^2 ( z_1(\bq) + z_2(\bq))}{\partial q_\mu \partial q_\nu} \Bigg|_{\bq=0}.   
\end{align}
To evaluate this, we need the expressions for $z_1(\bq)$ and $z_2(\bq)$. Let us first evaluate $z_1(\bq)$:
%\begin{widetext}
\begin{align}
\label{z1}
z_1(\bq) &=   -\int_0^\beta d\tau_1 \langle T_\tau H'(\bq,\tau_1) \rangle = -\beta \langle H'(\bq) \rangle \nonumber \\
&= -\beta \Big\{  \sum_{\mu\nu} q_\mu q_\nu \frac{N n_0}{2} \langle \phi_0 | \partial_\mu \partial_\nu \mathcal{H}(\bk_c) | \phi_0 \rangle +  \sum_\mu q_\mu \Big[ \sum_\bk \langle \delta c^\dag_\bk \partial_\mu \mathcal{H}(\bk) \delta c_\bk \rangle + \frac{1}{2}\sum_{\bk\nu} \langle \delta c^\dag_\bk \partial_\mu \partial_\nu \mathcal{H}(\bk) \delta c_\bk  \rangle q_\nu \Big]  \Big\} \nonumber \\
& = -\beta \Big\{  \sum_{\mu\nu}  \frac{ q_\mu q_\nu N n_0}{2} \langle \phi_0 | \partial_\mu \partial_\nu \mathcal{H}(\bk_c) | \phi_0 \rangle +  \sum_{\mu\nu}   \frac{q_\mu q_\nu}{2}\sum_\bk \langle \delta c^\dag_\bk \partial_\mu \partial_\nu \mathcal{H}(\bk) \delta c_\bk  \rangle    \Big\} .
\end{align}
%\end{widetext}
In the second line, the first (second) term arises from the condensate (fluctuation) Hamiltonian Eq.~\eqref{condterm} (Eq.~\eqref{flucterm}). The condensate-fluctuation Hamiltonian~\eqref{condflucterm} does not contribute as $\langle \delta c_{\bk\alpha}\rangle = 0$. The last line follows from the fact that in thermal equilibrium the current term vanishes, i.e. $\langle \delta c^\dag_\bk \partial_\mu \mathcal{H}(\bk) \delta c_\bk \rangle = 0$. 

Expression for $z_2(\bq)$ is more involved but can be obtained after a straightforward algebra. By expanding $z_2(\bq)$ up to second order in $\bq$, and realizing that $H_c(\bq)$ cannot give any contribution [as terms linear in $\bq$ are missing in Eq.~\eqref{condterm}] we obtain
%\begin{widetext}
\begin{align}
\label{z2}
&z_2(\bq) = \frac{1}{2} \int_0^\beta d\tau_1 \int_0^\beta d\tau_2 \langle T_\tau H'(\bq,\tau_1) H'(\bq,\tau_2) \rangle = \frac{1}{2}\int_0^\beta d\tau_1  \int_{0}^{\beta} d\tau \langle H'(\bq,\tau) H'(\bq,0) \rangle = \frac{\beta}{2} \int_{0}^{\beta} d\tau \langle H'(\bq,\tau) H'(\bq,0) \rangle \nonumber \\
&= \frac{\beta}{2} \int_{0}^{\beta} d\tau \sum_\mu \sum_\nu q_\mu q_\nu \Bigg[ \sum_{\bk,\bk',\alpha\beta\gamma\delta}\langle \delta c^\dag_{\bk\alpha}(\tau) \partial_\mu \mathcal{H}_{\alpha\beta} \delta c_{\bk\beta}(\tau) \delta c^\dag_{\bk'\gamma}\partial_\nu \mathcal{H}_{\gamma\delta}(\bk') \delta c_{\bk'\delta} \rangle \nonumber \\
& + n_0 N \Bigg \langle \Big(  \langle\phi_0 | \partial_\mu \mathcal{H}(\bk_c)  \delta c_{\bk_c}(\tau) +  h.c.  \Big)\Big(  \langle\phi_0 | \partial_\nu \mathcal{H}(\bk_c) \delta c_{\bk_c} +h.c.  \Big)\Bigg\rangle \Bigg].
\end{align}
%\end{widetext}
Here the first term in the final form arises from the fluctuations only and the second term comes from the condensate-fluctuation Hamiltonian. The notation of the last line, even though looking odd, is correct as $\langle \phi_0 |$ is a row vector of numbers and $\delta c_{\bk_c}$ is a column vector of operators $c_{\bk\alpha}$. By plugging the expressions~\eqref{z1} and \eqref{z2} in Eq.~\eqref{ddO}, we find
%\begin{widetext}
\begin{align}
\label{ddOdqq}
&D^s_{\text{old},\mu\nu} = \frac{\partial^2 \Omega(\bq)}{\partial q_\mu \partial q_\nu}\Bigg|_{\bq=0} =    N n_0 \langle \phi_0 | \partial_\mu \partial_\nu \mathcal{H}(\bk_c) | \phi_0 \rangle +  \sum_\bk \langle \delta c^\dag_\bk \partial_\mu \partial_\nu \mathcal{H}(\bk) \delta c_\bk  \rangle q_\nu   \nonumber \\
& - \int_{0}^{\beta} d\tau \Bigg[ \sum_{\bk,\bk',\alpha\beta\gamma\delta}\langle \delta c^\dag_{\bk\alpha}(\tau) \partial_\mu \mathcal{H}_{\alpha\beta} \delta c_{\bk\beta}(\tau) \delta c^\dag_{\bk'\gamma}\partial_\nu \mathcal{H}_{\gamma\delta}(\bk') \delta c_{\bk'\delta} \rangle \nonumber \\
& + n_0 N \Bigg\langle \Big(  \langle\phi_0 | \partial_\mu \mathcal{H}(\bk_c)  \delta c_{\bk_c}(\tau) +  h.c.  \Big)\Big(  \langle\phi_0 | \partial_\nu \mathcal{H}(\bk_c) \delta c_{\bk_c} +h.c.  \Big) \Bigg\rangle\Bigg].
\end{align}
%\end{widetext}
By comparing this form to the superfluid weight calculation presented in the Supplementary Material (SM) of Ref.~\cite{Julku2021b}, we can identify that the first line corresponds to the diamagnetic superfluid weight contribution, (i.e. Eqs.~(22) and (23) in the SM of Ref.~\cite{Julku2021b}), the second line corresponds to the fluctuation term (Eq.~(20) in the SM of Ref.~\cite{Julku2021b}) and the last line is the condensate-fluctuation superfluid contribution (Eq.~(17) in SM of Ref.~\cite{Julku2021b}). Thus, Eq.~\eqref{ddOdqq} corresponds to Eq.~(25) in SM of Ref.~\cite{Julku2021b}.

Following the notation of Ref.~\cite{Julku2021b}, we split $D^s_{\text{old}}$ to three terms, i.e. $D^s_{\text{old}} = D^s_1 + D^s_2 + D^s_3$, where $D^s_1$ is the pure condensate term, $D^s_2$ arises from the coupling between the condensate and the quantum fluctuations and $D^s_3$ is the pure quantum fluctuation contribution. For completeness, we provide the expressions for $D^s_1$, $D^s_2$ and $D^s_3$ derived in Ref.~\cite{Julku2021b}:
%\begin{widetext}
\begin{align}
&D^s_{1,\mu\nu} = n_0\partial_\mu \partial_\nu \epsilon_1(\bk_c) + n_0\sum_{n\neq 1}\Big\{[\epsilon_n(\bk_c) - \epsilon_0] \langle \partial_\mu \phi_0 | u_n(\bk_c)\rangle \langle u_n(\bk_c) | \partial_\nu \phi_0 \rangle + (\mu \leftrightarrow \nu) \Big\}, \\
&D^s_{2,\mu\nu} = -n_0  \lim_{\bq\rightarrow 0}  \sum_{ms}\frac{\langle \Phi_0 | \sigma_z \partial_\mu\mathcal{H}_B(\bk_c -\bq/2) | \psi^s_m(\bk_c-\bq)\rangle \langle \psi^s_m(\bk_c-\bq)|\sigma_z \partial_\nu \mathcal{H}_B(\bk_c -\bq/2)|\Phi_0\rangle }{E_m(\bk_c-s\bq)}, \\
&D^s_{3,\mu\nu} = \frac{1}{2N} \sum_\bk{}^{'} \sum_{mm'ss'} ss'\frac{n_B[sE_m(\bk_c+s\tilde{\bk})] - n_B[s'E_{m'}(\bk_c+s'\tilde{\bk})] }{s'E_{m'}(\bk_c+s'\tilde{\bk}) - sE_{m}(\bk_c+s\tilde{\bk})} \times \nonumber \\
&\Big[\langle \psi^{s'}_{m'}(\bk)| \partial_\mu \mathcal{H}_B(\bk) | \psi^{s}_{m}(\bk) \rangle \langle \psi^{s}_{m}(\bk)| \partial_\nu \mathcal{H}_B(\bk) | \psi^{s'}_{m'}(\bk) \rangle \nonumber \\
&-\langle \psi^{s'}_{m'}(\bk)| \sigma_z \partial_\mu \mathcal{H}_B(\bk) | \psi^{s}_{m}(\bk) \rangle \langle \psi^{s}_{m}(\bk)| \sigma_z \partial_\nu \mathcal{H}_B(\bk) | \psi^{s'}_{m'}(\bk) \rangle\Big] \label{Ds3},
\end{align}
%\end{widetext}
where $\tilde{\bk} \equiv \bk - \bk_c$. 

\section{Evaluating the correction term $D^s_{\text{corr}}$}\label{AppB}

%\begin{widetext}

To compute the superfluid weight correction term $D^s_{\text{corr}}$, given in Eq.~\eqref{int_res}, one needs to evaluate the partial derivatives $\frac{\partial^2 \Omega}{\partial q_\nu \partial \mu}$ and $\frac{\partial^2 \Omega}{\partial q_\nu \partial n_0}$. In this Appendix, we provide the expressions for these two quantities. To this end, we first note that one has
\begin{align}
&\frac{\partial \Omega}{\partial \mu} = \frac{\partial}{\partial \mu} \Bigg( -\frac{1}{\beta} \log Z \Bigg) = \frac{1}{Z} \text{Tr} \Big[ e^{-\beta H(\bq)} \frac{\partial H(\bq)}{\partial \mu}\Big] \nonumber \\
&= \Bigg\langle \frac{\partial H(\bq)}{\partial \mu} \Bigg\rangle
\end{align}
and similarly for $n_0$. From Eqs.~\eqref{origHc},~\eqref{origHf},~\eqref{condterm},~\eqref{condflucterm} and~\eqref{flucterm}, it can be easily seen that 
\begin{align}
\label{mu_part}
\frac{1}{N} \frac{\partial^2 \Omega}{\partial q_\nu \partial \mu} \Bigg|_{\bq=0} =\frac{1}{N}\frac{\partial}{\partial q_\nu}\Bigg\langle \frac{\partial H(\bq)}{\partial \mu} \Bigg\rangle \Bigg|_{\bq=0} = -\frac{\partial}{\partial q_\nu}n_{\text{tot}} \Bigg|_{\bq=0}    
\end{align}
and
\begin{align}
\label{n0_part}
& \frac{1}{N} \frac{\partial^2 \Omega}{\partial q_\nu \partial n_0} \Bigg|_{\bq=0} =\frac{1}{N}\frac{\partial}{\partial q_\nu}\Bigg\langle \frac{\partial H(\bq)}{\partial n_0} \Bigg\rangle \Bigg|_{\bq=0} = \frac{1}{N}\frac{\partial}{\partial q_\nu}\frac{\partial H_c}{\partial n_0} \Bigg|_{\bq=0} + \frac{1}{N} \frac{\partial}{\partial q_\nu} \Bigg\langle \frac{\partial H_f}{\partial n_0} \Bigg\rangle \Bigg|_{\bq=0} \nonumber \\
& = \frac{1}{N}\frac{\partial}{\partial q_\nu}\frac{\partial H_c}{\partial n_0} \Bigg|_{\bq=0} + \frac{\partial}{\partial q_\nu} \frac{U}{2N} \sum_{\bk \alpha} \Big[ \big(\langle \alpha | \phi_0 \rangle^2 \langle \delta c_{\bk\alpha}^\dag \delta c^\dag_{2\bk_c-\bk \alpha} \rangle + \text{h.c.}\big) +4|\langle \alpha | \phi_0 \rangle|^2 \langle \delta c_{\bk\alpha}^\dag c_{\bk\alpha} \rangle \Big] \Bigg|_{\bq=0}.
\end{align}
Here $H_c$ is a number given by the first form of the second line in Eq.~\eqref{origHc} and the last term in Eq.~\eqref{n0_part} arises from the fluctuation Hamiltonian given by Eq.~\eqref{origHf}. Note that all the contributions from the linear term Eq.~\eqref{condflucterm} vanish as $\langle \delta c_{\bk\alpha} \rangle = 0$.

To evaluate Eqs.~\eqref{mu_part} and~\eqref{n0_part}, we expand the expectation values in terms of small $\bq$ up to linear order, in the same way as earlier in case of the grand canonical potential in Eq.~\eqref{gcp}. We demonstrate the calculation for Eq.~\eqref{mu_part}. We start by writing
\begin{align}
\label{ntot}
&n_{\text{tot}} = \frac{1}{Z(\bq)}\Tr[e^{-\beta H(\bq)} n] = \frac{1}{Z(\bq)}\Tr[e^{-\beta H}\tilde{U}(\beta,0) n] \approx \frac{Z(0)}{Z(0)[1 + z_1(\bq)]}\Bigg[ n_{\text{tot}}(\bq=0) - \int_0^\beta d\tau \langle T_\tau H'(\bq,\tau) n  \rangle_0 \Bigg] \nonumber \\
&= \frac{1}{1 + z_1(\bq)}\Bigg[ n_{\text{tot}}(\bq=0) -\int_0^\beta d\tau \langle T_\tau H'(\bq,\tau) n  \rangle_0 \Bigg]
\end{align}
where $n$ is the total density operator, $n_{\text{tot}}(\bq=0)$ is the total density at $\bq=0$ and the notation $\langle \dots \rangle_0$ implies the expectation value taken with respect to the equilibrium state at $\bq=0$. In this case, as $n$ is quadratic in the fluctuation operators, we take here $H'(\bq) = \sum_\mu q_\mu \sum_\bk \delta c^\dag_\bk \partial_\mu \mathcal{H}(\bk) \delta c_\bk$ as is evident from Eq.~\eqref{flucterm} (higher terms in $\bq$ vanish when taking the \bq-derivative and subsequently setting $\bq = 0$). From Eq.~\eqref{ntot} we can now obtain the following
%\begin{align}
%&\frac{ \partial n_{\text{tot}}}{\partial q_\nu} \Bigg|_{\bq=0} = \beta\sum_\bk \langle \delta c^\dag_{\bk} \partial_\nu \mathcal{H}(\bk) \delta c_{\bk}\rangle \langle T_\tau H'(\bq) n\rangle - \int_0^\beta d\tau \langle T_\tau \sum_{\bk} \delta c^\dag_\bk(\tau) \partial_\nu \mathcal{H}(\bk) \delta c_\bk(\tau) n \rangle \nonumber \\
%& = - \int_0^\beta d\tau \langle T_\tau \sum_{\bk} \delta c^\dag_\bk(\tau) \partial_\nu \mathcal{H}(\bk) \delta c_\bk(\tau) n \rangle
%\end{align}
%where the first term vanishes at $\bq =0$ (or equivalently due to the vanishing current expectation value) and we have dropped subscript $0$ for simplicity. By recalling that $n = n_0 + \frac{1}{N} \sum_{\bk\alpha} \delta c_{\bk\alpha}^\dag c_{\bk\alpha}$, we note that 
\begin{align}
&\frac{ \partial n_{\text{tot}}}{\partial q_\nu} \Bigg|_{\bq=0} = - \int_0^\beta d\tau \langle T_\tau \sum_{\bk} \delta c^\dag_\bk(\tau) \partial_\nu \mathcal{H}(\bk) \delta c_\bk(\tau) n \rangle \nonumber \\
\end{align}
where we have dropped subscript $0$ for simplicity. By recalling that $n = n_0 + \frac{1}{N} \sum_{\bk\alpha} \delta c_{\bk\alpha}^\dag c_{\bk\alpha}$, we note that 
\begin{align}
\label{ntot2}
& \frac{ \partial n_{\text{tot}}}{\partial q_\nu} \Bigg|_{\bq=0} = - \int_0^\beta d\tau \Bigg\langle T_\tau \sum_{\bk} \delta c^\dag_\bk(\tau) \partial_\nu \mathcal{H}(\bk) \delta c_\bk(\tau) \frac{1}{N}\sum_{\bk'\alpha} \delta c_{\bk\alpha}^\dag \delta c_{\bk\alpha} \Bigg\rangle  
\end{align}
as the expectation value of the current operator vanishes in equilibrium, i.e.~$\sum_\bk \langle \delta c_\bk^\dag \partial_\nu\mathcal{H}(\bk) \delta c_\bk \rangle = 0$. 

By now employing the standard Green's function techniques, as was done e.g. in Ref.~\cite{Julku2021b}, one can show that Eq.~\eqref{ntot2} at zero temperature reduces to 
\begin{align}
\label{mu_res}
\frac{1}{N} \frac{\partial^2 \Omega}{\partial q_\nu \partial \mu} \Bigg|_{\bq=0} = -\frac{ \partial n_{\text{tot}}}{\partial q_\nu} \Bigg|_{\bq=0} = \frac{1}{N} \sum_{\bk m m'} \frac{\Re[\langle \psi^+_m(\bk) | \partial_\nu L_\bk| \psi^-_{m'}(\bk) \rangle ]}{E_m(\bk) + E_{m'}(2\bk_c-\bk)}.
\end{align}
In the same way, one can show that at zero temperature one has for a uniform condensate
\begin{align}
\label{n0_res}
&\frac{1}{N} \frac{\partial^2 \Omega}{\partial q_\nu \partial n_0} \Bigg|_{\bq=0} = -\frac{U}{N} \sum_{\bk mm'} \frac{\Re[\mathcal{A}_{mm'}(\bk) \langle \psi^-_{m'}(\bk) | \partial_\nu L(\bk)| \psi^+_m(\bk) \rangle]}{E_m(\bk) + E_{m'}(2\bk_c-\bk)} - \frac{2U}{MN}\sum_{\bk m m'} \frac{\Re[\langle \psi^+_m(\bk) | \partial_\nu L_\bk| \psi^-_{m'}(\bk) \rangle ]}{E_m(\bk) + E_{m'}(2\bk_c-\bk)}
\end{align}
with
\begin{align}
& \mathcal{A}_{mm'}(\bk) = \langle \psi^+_m(\bk) | A | \psi^-_{m'}(\bk) \rangle  \\
& A = \begin{bmatrix}
 0 & \phi_0 \\ \phi^*_0 & 0
\end{bmatrix}  \\
& [\phi_0]_{\alpha\beta} = \delta_{\alpha,\beta} \langle \alpha | \phi_0 \rangle^2.
\end{align}

With Eqs.~\eqref{mu_res} and~\eqref{n0_res} one can compute the required partial derivatives with respect to $\bq$ without the knowledge of the state at finite $\bq$. However, as can be seen from Eq.~\eqref{int_res}, evaluating $D^s_{\text{corr}}$ requires also computing the full derivatives $\frac{d n_0}{d q_\mu}$ and $\frac{d \mu}{d q_\mu}$. This can be done numerically by solving the Bogoliubov problem at fixed total density for small finite $\bq$ in the vicinity of $\bq=0$ such that $|\bq| \ll \Delta k \sim 2\pi/\sqrt{N}$. Then the Bogoliubov theory boils down to solving the quadratic Hamiltonian of the form
\begin{align}\label{BogMatrix}
&\mathcal{H}_B(\bk) = \begin{bmatrix}
	\mathcal{H}(\bk + \bq) -\mu_{\textrm{eff}} &  \Delta \\
	\Delta^* & \mathcal{H}^*(\bq - \bk) -\mu_{\textrm{eff}},
\end{bmatrix}, \nonumber \\
& \Psi_\bk = [c_{\bk + \bq 1}, c_{\bk + \bq 2},...,c_{\bk + \bq M}, c^\dag_{\bq - \bk 1},...,c^\dag_{\bq - \bk M}]^T.
\end{align}
By computing $n_0$ and $\mu = \epsilon_0 + \frac{Un_0}{M}$ as a function of small $\bq$, one can then determine $\frac{d n_0}{d q_\mu}$ and $\frac{d \mu}{d q_\mu}$.

%\end{widetext}
%\begin{widetext}
\section{Second order perturbation theory for the speed of sound}\label{AppB}

To obtain the second order perturbation result for the speed of sound, one needs to first write down the full Bogoliubov matrix $L(\bk)$ in the Bloch basis by performing the basis transformation  $\mathcal{U}(\bk)$ as
\begin{align}
\mathcal{U}(\bk) = \begin{bmatrix}
 U(\bk) & 0 \\ 0 & U^*(2\bk_c -\bk).
\end{bmatrix}    
\end{align}
By transforming $L(\bk) \rightarrow \mathcal{U}^\dagger(\bk)L(\bk)\mathcal{U}(\bk)$ and writing the momentum as $\bk = \bk_c + \bq$, $L(\bk)$ can be written in the Bloch basis as
%\begin{widetext}
\begin{align}
\label{Lk}
L(\bk) =  \begin{bmatrix}
	D(\bk_c + \bq) & U^\dag(\bk_c + \bq)\Delta U^*(\bk_c - \bq) \\
	-U^T(\bk_c - \bq)\Delta^* U(\bk_c + \bq) & -D(\bk_c - \bq)
\end{bmatrix}.
\end{align}
%\end{widetext}
In case of the uniform BEC, $D(\bk)$ is a diagonal matrix and reads $[D(\bk_c + \bq)]_{ij} = \delta_{ij}( \epsilon_j(\bk_c + \bq) - \epsilon_0 + \tilde{U} )$ with $\tilde{U} = \frac{Un_0}{M}$. By only keeping the lowest Bloch band and projecting out all the other Bloch bands in Eq.~\eqref{Lk}, one gets a $2\times 2$ projected Bogoliubov matrix $L_p(\bk)$, i.e. Eq.~\eqref{lk2}. By diagonalizing $L_p$, one obtains 
%\begin{widetext}
\begin{align}
&L_p(\bk) | \psi^{\pm}_1 (\bk) \rangle = \pm E_1(\bk) | \psi^{\pm}_1 (\bk) \rangle \\
& E_1(\bk_c + \bq) = \frac{Un_0}{M}\tilde{D}(\bq) \equiv E^{(0)}_\bq \\
& | \psi^{+}_1 (\bk_c + \bq) \rangle = \frac{\tilde{U}|\alpha(\bq)|}{\sqrt{2}\sqrt{E_1(\bk_c + \bq)^2 + \tilde{U} E_1(\bk_c + \bq)}}
\begin{bmatrix}
 -\frac{\tilde{U} + E_1(\bk_c + \bq)}{\tilde{U} \alpha^*(\bq)} \\
 1
\end{bmatrix} \equiv \begin{bmatrix}
u_1(\bq) \\
v_1(\bq)
\end{bmatrix}
\end{align}
%\end{widetext}
The effect of other Bloch bands can be taken into account perturbatively by writing Eq.~\eqref{Lk} as $L(\bk) = L_0(\bk) + L'(\bk)$, where $L_0(\bk)$ reads (after rearranging the indices accordingly)
\begin{align}
&L_0(\bk) = \begin{bmatrix}
D^+(\bk) & 0 & 0 \\
0 & L_p(\bk) & 0 \\
0 & 0 & D^-(\bk)
\end{bmatrix},   \quad \text{with} \\
& [D^{\pm}(\bk)]_{ij} = \pm [\epsilon_{i+1}(\bk_c \pm \bq) - \epsilon_0 + \tilde{U}] \delta_{ij}.
\end{align}
Here matrices $D^{\pm}(\bk)$ are $(M-1)\times (M-1)$ matrices. By now applying the standard perturbation theory~\cite{Liboff} for $L(\bk)$ with the perturbation term being $L'(\bk)$, one can write the Goldstone energy as a perturbative series up to the second order as
\begin{align}
&E_1(\bk_c + \bq) = E^{(0)}_\bq  + E^{(1)}_\bq  + E^{(2)}_\bq \nonumber \\
& = E^{(0)}_\bq  + E^{(2)}_\bq.
\end{align}
In the second line we have used the fact that for our choice of $L_0(\bk)$ and $L'(\bk)$ the first-order term vanishes, i.e. $E^{(1)}_\bq = 0$. The second order contribution reads
\begin{align}
&E^{(2)}_\bq = \sum_{j=2}^M \frac{|v_1(\bq) \langle u_1(\bk_c-\bq) |\Delta| u^*_j(\bk_c+\bq) \rangle|^2}{E_1(\bk_c + \bq) - [D^+(\bk)]_{j-1,j-1}} \nonumber \\
& -\sum_{j=2}^M \frac{|u_1(\bq) \langle u_1(\bk_c+\bq) |\Delta| u^*_j(\bk_c-\bq) \rangle|^2}{E_1(\bk_c + \bq) - [D^-(\bk)]_{j-1,j-1}}.
\end{align}
Here the first (second) line corresponds to the coupling of $L_p$ to the particle (hole) sector of $L_0(\bk)$, i.e. to $D^+$ ($D^-$). It is straightforward to show that in the limit of $\bq \rightarrow 0$ the coherence factors of the flat band BEC approach the same limit, i.e.~$u_1(\bq) \rightarrow v_1(\bq) \rightarrow \frac{\tilde{U}}{2E_\bq^{(0)}}$. Moreover, $|E_1(\bk_c + \bq)| \ll |[D^{\pm}(\bk)]_{j-1,j-1}|$ for $\bq \rightarrow 0$. Therefore, for $\bq \rightarrow 0$ we can write
\begin{align}
 &E^{(2)}_\bq = -\sum_{j=2}^M \frac{\tilde{U}|\langle u_1(\bk_c-\bq) |\Delta| u^*_j(\bk_c+\bq) \rangle|^2}{2 E^{(0)}_\bq[ \epsilon_{j}(\bk_c + \bq) - \epsilon_0 + \tilde{U} ]}   \nonumber \\
 & -\sum_{j=2}^M \frac{\tilde{U}|\langle u_1(\bk_c+\bq) |\Delta| u^*_j(\bk_c-\bq) \rangle|^2}{2 E^{(0)}_\bq[ \epsilon_{j}(\bk_c - \bq) - \epsilon_0 + \tilde{U} ]}.
\end{align}
This is the form used in Sec.~\ref{Sec_cs}.
\end{widetext}


%\bibliography{bib_file,bib_tbg,biblio_Nat_Phys_Rev_Sebastiano}

\end{document}